\documentclass[a4paper,12pt]{article}

\usepackage{amsmath,amsthm,amssymb}
\usepackage[dvips]{graphicx}
\usepackage{multirow}
\usepackage{here}
\usepackage{bm}
\usepackage{float}
\usepackage{latexsym}
\usepackage{subfigure}
\usepackage{booktabs}
\usepackage{lscape}
\usepackage{color}

\DeclareMathOperator*{\argmin}{arg\,min}

\begin{document}
\setlength{\baselineskip}{18.5pt}
\author{Toshihiro Hirano\thanks{Graduate School of Economics, The University of Tokyo, Hongo 7-3-1, 
Bunkyo-ku, Tokyo 113-0033, 
Japan, Currently at NEC Corporation, 1753, Shimonumabe, Nakahara-ku, Kawasaki, Kanagawa 211-8666, Japan. E-mail: nennekog@hotmail.co.jp}
}
\title{Modified Linear Projection for Large Spatial Data Sets}
\maketitle

\begin{abstract}

Recent developments in engineering techniques for spatial data collection such as geographic information systems have resulted in an increasing need for methods to analyze large spatial data sets. These sorts of data sets can be found in various fields of the natural and social sciences. However, model fitting and spatial prediction using these  large spatial data sets are impractically time-consuming, because of the necessary matrix inversions. Various methods have been developed to deal with this problem, including a reduced rank approach and a sparse matrix approximation. In this paper, we propose a modification to an existing reduced rank approach to capture both the large- and small-scale spatial variations effectively. We have used simulated examples and an empirical data analysis to demonstrate that our proposed approach consistently performs well when compared with other methods. In particular, the performance of our new method does not depend on the dependence properties of the spatial covariance functions.
\end{abstract}


\section{Introduction}
\label{intro}

Spatial data set analysis has been attracting an increasing amount of attention from various fields such as environmental science and economics, but is often impractical for large spatial data sets. This is because model fitting and spatial prediction in a Gaussian process model involve the inversion of an $n \times n$ covariance matrix for a data set of size $n$, which typically requires $O(n^3)$ operations. There is a rich literature regarding  efficient computation for large spatial data sets (e.g., Stein et al. 2004; Fuentes 2007; Matsuda and Yajima 2009; Lindgren et al. 2011). 

In this paper, we consider two recently developed approaches that appeal as general purpose methodologies. The first approach is based on a reduced rank approximation of the underlying process. Cressie and Johannesson (2008) considered fixed rank approaches for kriging in large spatial data sets. Banerjee et al. (2008) proposed a predictive process that used a finite number of knots and Finley et al. (2009) corrected a bias in the predictive process. Recently, Banerjee et al. (2013) developed a linear projection approach in the literature of Gaussian process regression (see, e.g., Rasmussen and Williams 2006). As in Banerjee et al. (2013), this is an extension of the predictive process and has the advantage of avoiding the complicated knot selection problem.

The second approach is covariance tapering proposed by Furrer et al. (2006). The basic idea of the covariance tapering is to reduce a spatial covariance function to zero beyond some range by multiplying the true spatial covariance function by a positive definite, compactly supported function. Then, the resulting covariance matrix is sufficiently sparse to achieve computational efficiency with the sparse matrix algorithm (see, e.g., Davis 2006). Furrer et al. (2006) proved the asymptotic efficiency of the best linear unbiased predictor (BLUP) using the covariance tapering which is called the tapered BLUP for the original BLUP. Kaufman et al. (2008) applied the covariance tapering to the log-likelihood function and showed that the estimators maximizing the tapered approximation of the log-likelihood are strongly consistent. Hirano and Yajima (2013) investigated the asymptotic property of the prediction by the covariance tapering in a transformed random field.

Sang and Huang (2012) demonstrated that the predictive process fails to accurately approximate the small-scale dependence structure and the covariance tapering fails at large-scales. They proposed a combination of the predictive process and the covariance tapering, which is called a full scale approximation. Our paper confirms the same approximation property for the linear projection as the predictive process through some examples. We will show only one example in this paper. To deal with this problem, we propose a modified linear projection using the covariance tapering based on the work of Sang and Huang (2012).

The main contributions of this paper are to propose a linear projection using a modification by the compactly supported correlation function and investigate theoretical justification. Furthermore, we have used simulated examples and an empirical analysis based on the air dose rate data to demonstrate that our proposed method works well when compared with the linear projection and the covariance tapering, regardless of the strength of spatial correlation and nonstationarity. Our work can be regarded as an extension of Banerjee et al. (2008), Finley et al. (2009), Sang and Huang (2012), and Banerjee et al. (2013). However, we have only focused on the linear projection approach proposed by Banerjee et al. (2013) for comparison purposes because it outperformed the predictive process in the simulations and empirical studies of Banerjee et al. (2013) and the methods proposed by Finley et al. (2009) and Sang and Huang (2012) are a modification of the predictive process by the indicator function and the compactly supported correlation function respectively. Moreover, the selection of tuning parameters and many indices such as the accuracy of the  estimation and prediction and computational time make fair comparisons difficult.

The remainder of this paper is organized as follows. We introduce a linear regression model and Bayesian analysis for spatial data sets in Section 2. In Section 3, we review the linear projection approach and its algorithm. Section 4 presents our proposed modified linear projection. In Section 5, we present the results of computer experiments that  compared the performance of our method with that of the linear projection and covariance tapering. Section 6 provides an  empirical analysis based on the air dose rate in Chiba prefecture of eastern Japan. Our conclusions and future studies are discussed in Section 7. Technical proofs of the propositions are given in the Appendix. 

\section{Linear regression model and Bayesian analysis for spatial data sets}




For $\bm{s} = (s_1,\ldots,s_d)^{'} (\in D \subset \mathbb{R}^d)$, consider the linear regression model of the form
\begin{align}
Y(\bm{s}) = \bm{x}(\bm{s})^{'} \bm{\beta} + W(\bm{s}) + \epsilon(\bm{s}),
\label{model}
\end{align}
where $Y(\bm{s})$ is a dependent variable at a location $\bm{s}$, $\bm{x}(\bm{s}) = (x_1(\bm{s}), \ldots , x_p(\bm{s}))^{'}$ is a $p$-vector of nonstochastic regressors, $\bm{\beta} = (\beta_1, \ldots, \beta_p)^{'}$ is a vector of unknown regression coefficients, and the prime denotes the transposition. The residual of this regression is decomposed into a zero-mean Gaussian process $W(\bm{s})$ with a valid covariance function $C_{W}(\bm{s},\bm{s}^{*}) = \mbox{cov}(W(\bm{s}), W(\bm{s}^{*}) )$ ($\bm{s}$, $\bm{s}^{*} \in D$) and $\epsilon(\bm{s})$ which is a zero-mean independent process following a normal distribution with a variance $\tau^2$ for any location $\bm{s}$. $\epsilon(\bm{s})$ represents the possibility of measurement error and/or microscale variability and is often referred to as a nugget effect (see, e.g., Cressie 1993). It is assumed that $\{ W(\bm{s}) \}$ and $\{ \epsilon(\bm{s}) \}$ are independent. We specify that $C_{W}(\bm{s},\bm{s}^{*}) = \sigma^2 \rho_W (\bm{s},\bm{s}^{*};\bm{\theta})$ where $\sigma^2 = \mbox{var}(W(\bm{s}))$, $\rho_W$ is a correlation function of the spatial process $W(\bm{s})$, and $\bm{\theta}$ is a vector of correlation parameters. 

Along with a $p \times 1$ vector of spatially referenced regressors $\bm{x}(\bm{s})$, we observe the dependent variable $Y(\bm{s})$ at given sampling locations $\bm{s}_1,\ldots,\bm{s}_n \in D$. Denote $\bm{Y}=(Y(\bm{s}_1),\ldots,Y(\bm{s}_n))^{\prime}$ and $\Omega = (\bm{\beta}, \tau^2, \sigma^2, \bm{\theta})$. Then, the probability density function of $\bm{Y}$ is
\begin{align*}
 f(\bm{Y} | \Omega) = (2 \pi)^{- \frac{n}{2}}  \left| \Sigma_W + \tau^2 \bm{I} \right|^{- \frac{1}{2}} \exp \left\{  - \frac{1}{2} (\bm{Y} - X \bm{\beta})^{\prime} \left( \Sigma_W + \tau^2 \bm{I} \right)^{-1} (\bm{Y} - X \bm{\beta}) \right\},
\label{prob} 
\end{align*}
where $(\Sigma_W)_{ij} = \sigma^2 \rho_W (\bm{s}_i,\bm{s}_j;\bm{\theta})$ $(i,j=1,\ldots,n)$, $\bm{I}$ is an $n \times n$ identity matrix, and $X = (\bm{x}(\bm{s}_1),\ldots,\bm{x}(\bm{s}_n))^{\prime}$. The goal is to estimate the parameters $\Omega = (\bm{\beta}, \tau^2, \sigma^2, \bm{\theta})$ and predict $Y(\bm{s}_0)$ at an unobserved location $\bm{s}_0 \in D$ based on $\bm{Y}$. Note that $\bm{x}(\bm{s}_0)$ is observed.

In this paper, we take a Bayesian approach and use a simulation method, namely, the Markov chain Monte Carlo (MCMC) method to generate samples from the posterior distribution and conduct the statistical inference with respect to the model parameters. The Bayesian approach assigns prior distributions to $\Omega = (\bm{\beta}, \tau^2, \sigma^2, \bm{\theta})$ and the MCMC method is used to draw samples of the model parameters from the posterior distribution
\[
\pi(\Omega | \bm{Y}) \propto f(\bm{Y} | \Omega) \pi(\bm{\beta}) \pi(\tau^2) \pi(\sigma^2) \pi(\bm{\theta}).
\]
For prior distributions of $\bm{\beta}$, $\sigma^2$, and $\tau^2$, we assume
\begin{align}
\bm{\beta} \sim {\cal N }(\bm{\mu}_{\bm{\beta}}, \Sigma_{\bm{\beta}}), \quad \tau^2 \sim {\cal IG }(a_1,b_1), \quad \sigma^2 \sim {\cal IG }(a_2,b_2),
\label{prior}
\end{align}
where ${\cal N }(\bm{\mu}_{\bm{\beta}}, \Sigma_{\bm{\beta}})$ and ${\cal IG }(a_i,b_i)$ ($i=1,2$) respectively denote the multivariate normal distribution and inverse gamma distributions with probability density functions
\begin{align*}
&\pi(\bm{\beta}) \propto \left| \Sigma_{\bm{\beta}} \right|^{- \frac{1}{2}} \exp \left\{ - \frac{1}{2} (\bm{\beta} - \bm{\mu}_{\bm{\beta}})^{\prime} \Sigma_{\bm{\beta}}^{-1} (\bm{\beta} - \bm{\mu}_{\bm{\beta}}) \right\}, 
\pi(\tau^2) \propto (\tau^2)^{-(a_1 + 1)} \exp \left( - \frac{b_1}{\tau^2} \right), \\
&\pi(\sigma^2) \propto (\sigma^2)^{-(a_2 + 1)} \exp \left( - \frac{b_2}{\sigma^2} \right).
\end{align*}
Since the prior specifications for $\bm{\theta}$ will depend on the choice of the correlation function $\rho_W (\bm{s},\bm{s}^{*};\bm{\theta})$, details of the estimation are discussed in Sections 5 and 6. We implement the MCMC algorithm in four stages:
\begin{enumerate}
\item Generate $\bm{\beta}|\tau^2,\sigma^2,\bm{\theta},\bm{Y}$.
\item Generate $\tau^2|\bm{\beta},\sigma^2,\bm{\theta},\bm{Y}$.
\item Generate $\sigma^2|\bm{\beta},\tau^2,\bm{\theta},\bm{Y}$.
\item Generate $\bm{\theta}|\bm{\beta},\tau^2,\sigma^2,\bm{Y}$.
\end{enumerate}
Note that $\Sigma_W$ depends on $\sigma^2$ and $\bm{\theta}$.

\noindent
\textit{Generation of $\bm{\beta}$.}

The conditional posterior probability density function of $\bm{\beta}$ is
\[
\pi(\bm{\beta}|\tau^2,\sigma^2,\bm{\theta},\bm{Y}) \sim {\cal N }(\bm{\mu}_{\bm{\beta}|\cdot},\Sigma_{\bm{\beta}|\cdot}),
\]
where 
\begin{align*}
\Sigma_{\bm{\beta}|\cdot} = \left\{ \Sigma_{\bm{\beta}}^{-1} + X^{\prime} \left( \Sigma_W + \tau^2 \bm{I} \right)^{-1} X \right\}^{-1} \intertext{and} \bm{\mu}_{\bm{\beta}|\cdot} = \Sigma_{\bm{\beta}|\cdot} \left\{ \Sigma_{\bm{\beta}}^{-1}  \bm{\mu}_{\bm{\beta}} + X^{\prime} \left( \Sigma_W + \tau^2 \bm{I} \right)^{-1} \bm{Y}  \right\}.
\end{align*}

$\tau^2$ and $\sigma^2$ are updated using Metropolis steps (see, e.g., Gelman et al. 2004). Random-walk Metropolis steps with normal proposals are typically adopted.

\noindent
\textit{Generation of $\tau^2$.}

Given the current value $\tau^2$, propose a candidate ${\tau^2}^{*} = \tau^2 + z_1$, $z_1 \sim {\cal N }(0, \sigma_1^2)$ and accept it with probability
\[
\min \left[ \frac{   \left| \Sigma_W + {\tau^2}^{*} \bm{I} \right|^{- \frac{1}{2}} \exp \left\{  - \frac{1}{2} (\bm{Y} - X \bm{\beta})^{\prime} \left( \Sigma_W + {\tau^2}^{*} \bm{I} \right)^{-1} (\bm{Y} - X \bm{\beta}) - \frac{b_1}{{\tau^2}^{*}}  \right\} {({\tau^2}^{*})}^{-(a_1 +1)}     }{  \left| \Sigma_W + \tau^2 \bm{I} \right|^{- \frac{1}{2}} \exp \left\{  - \frac{1}{2} (\bm{Y} - X \bm{\beta})^{\prime} \left( \Sigma_W + \tau^2 \bm{I} \right)^{-1} (\bm{Y} - X \bm{\beta}) - \frac{b_1}{\tau^2}  \right\} {(\tau^2)}^{-(a_1 +1)}   } , 1  \right].
\]

\noindent
\textit{Generation of $\sigma^2$.}

Given the current value $\sigma^2$, propose a candidate ${\sigma^2}^{*} = \sigma^2 + z_2$, $z_2 \sim {\cal N }(0, \sigma_2^2)$ and accept it with probability
\[
\min \left[ \frac{   \left| \Sigma_W^{*} + \tau^2 \bm{I} \right|^{- \frac{1}{2}} \exp \left\{  - \frac{1}{2} (\bm{Y} - X \bm{\beta})^{\prime} \left( \Sigma_W^{*} + \tau^2 \bm{I} \right)^{-1} (\bm{Y} - X \bm{\beta}) - \frac{b_2}{{\sigma^2}^{*}}  \right\} {({\sigma^2}^{*})}^{-(a_2 +1)}     }{  \left| \Sigma_W + \tau^2 \bm{I} \right|^{- \frac{1}{2}} \exp \left\{  - \frac{1}{2} (\bm{Y} - X \bm{\beta})^{\prime} \left( \Sigma_W + \tau^2 \bm{I} \right)^{-1} (\bm{Y} - X \bm{\beta}) - \frac{b_2}{\sigma^2}  \right\} {(\sigma^2)}^{-(a_2 +1)}   } , 1  \right],
\]
where $(\Sigma_W^{*})_{ij} = {\sigma^2}^{*} \rho_W (\bm{s}_i,\bm{s}_j;\bm{\theta})$ $(i,j=1,\ldots,n)$. The tuning parameters $\sigma_1^2$ and $\sigma_2^2$ are chosen such that the average of the acceptance rates in each iteration is approximately 40\%. As we have previously mentioned, the generation of $\bm{\theta}$ will be discussed in subsequent sections.

It is computationally expensive to calculate the determinant and inverse of the $n \times n$ matrix $\Sigma_W + \tau^2 \bm{I}$ for large spatial data sets. In particular, the inverse matrix calculation requires $O(n^3)$ operations. For each sampling procedure, we must calculate the determinant and inverse of the $n \times n$ matrix $\Sigma_W + \tau^2 \bm{I}$. Thus, the computational complexity of the above MCMC algorithm is challenging for large spatial data sets because a large number of samples from the posterior distribution are usually needed.

The Bayesian prediction is to obtain the predictive distribution
\[
\pi(Y(\bm{s}_0) | \bm{Y}) = \int \pi(Y(\bm{s}_0) | \bm{Y},\Omega) \pi(\Omega | \bm{Y}) d \Omega. 
\]
For a given $\Omega$, 
\begin{align*}
\pi(Y(\bm{s}_0) | \bm{Y},\Omega) \sim {\cal N } \left(\bm{x}(\bm{s}_0)^{\prime}\bm{\beta} + \bm{c}_{Y,\bm{s}_0}^{\prime} \left( \Sigma_W + \tau^2 \bm{I} \right)^{-1} \left(\bm{Y} - X \bm{\beta} \right), \sigma^2 + \tau^2 - \bm{c}_{Y,\bm{s}_0}^{\prime} \left( \Sigma_W + \tau^2 \bm{I} \right)^{-1} \bm{c}_{Y,\bm{s}_0} \right),
\end{align*}
where $\bm{c}_{Y,\bm{s}_0} = \left( \mbox{cov}(Y(\bm{s}_0), Y(\bm{s}_1)), \ldots, \mbox{cov}(Y(\bm{s}_0), Y(\bm{s}_n) ) \right)^{\prime}$. The predictive distribution is sampled by composition, drawing $Y^{(l)}(\bm{s}_0) \sim \pi(Y(\bm{s}_0) | \bm{Y},\Omega^{(l)})$ for each $\Omega^{(l)}$ ($l = 1,\ldots,L$) where $\Omega^{(l)}$ is the $l$th sample from the posterior distribution $\pi(\Omega | \bm{Y})$ and $L$ is the total number of samples given in the MCMC algorithm. The mean squared prediction error (MSPE) is computed using
\[
\frac{1}{M}\sum_{m=1}^{M} \left( Y(\bm{s}_{0,m}) - \frac{1}{L} \sum_{l=1}^{L} Y^{(l)}(\bm{s}_{0,m}) \right)^2,
\]
where $Y(\bm{s}_{0,m})$ ($m = 1, \ldots, M$) and $\sum_{l=1}^{L} Y^{(l)}(\bm{s}_{0,m})/L$ denote the test data sets and the sample analogue of the mean of the predictive distribution respectively. Since sampling from $\pi(Y(\bm{s}_0) | \bm{Y},\Omega^{(l)})$ also involves the inverse of the $n \times n$ matrix $\Sigma_W + \tau^2 \bm{I}$, the computation becomes a more formidable one for large spatial data sets.

Finally, we compare some existing approximation methods using the deviance information criterion (DIC) (Spiegelhalter et al. 2002). It is used as a Bayesian measure of fit or adequacy and is defined as
\[
DIC = E_{\Omega | \bm{Y}}[ D(\Omega) ] + p_D,
\]
where $D(\Omega) = -2 \log f(\bm{Y} | \Omega)$ and $p_D = E_{\Omega | \bm{Y}}[ D(\Omega) ] - D(E_{\Omega | \bm{Y}}[ \Omega ])$. $E_{\Omega | \bm{Y}}[ \cdot ] $ represents the expectation under the posterior distribution $\pi(\Omega | \bm{Y})$. To compute $E_{\Omega | \bm{Y}}[ D(\Omega) ]$ and $E_{\Omega | \bm{Y}}[ \Omega ]$, we use the sample analogues
\begin{align*}
\frac{1}{L} \sum_{l = 1}^{L} D(\Omega^{(l)}) \quad \text{and} \quad \frac{1}{L} \sum_{l = 1}^{L} \Omega^{(l)}.
\end{align*}
DIC also includes the inversion of the $n \times n$ matrix. 

\section{Linear projection approach}

In this section, we review the linear projection approach proposed by Banerjee et al. (2013). This method was developed to efficiently compute Gaussian process regression. However, it can be applied to the efficient computation of the Bayesian analysis for large spatial data sets and the linear projection approach is regarded as an extension of predictive process models (Banerjee et al. 2013).

As a first step, for $\bm{s} \in D$, define
\begin{align*}
W_{approx}(\bm{s}) = E[W(\bm{s})|\Phi \bm{W}] = \bm{c}_{W,\bm{s}}' \Phi' (\Phi \Sigma_{W} \Phi')^{-1} \Phi \bm{W},
\end{align*}
where $\bm{W} = (W(\bm{s}_1),\ldots,W(\bm{s}_n))^{\prime}$, $\bm{c}_{W,\bm{s}} = \left( \mbox{cov}(W(\bm{s}), W(\bm{s}_1)), \ldots, \mbox{cov}(W(\bm{s}), W(\bm{s}_n) ) \right)^{\prime} = \left( C_{W}(\bm{s}, \bm{s}_1), \ldots, C_{W}(\bm{s}, \bm{s}_n) \right)^{\prime}$, and $\Phi$ is an $m \times n$ matrix with full row-rank ($m \le n$) and a row-norm equal to unity to avoid scale problems. In this case, for $\bm{s}, \bm{s}^{*} \in D$,
\begin{align*}
C_{approx}(\bm{s},\bm{s}^{*}) = \mbox{cov}(W_{approx}(\bm{s}), W_{approx}(\bm{s}^{*})) = \bm{c}_{W,\bm{s}}^{\prime} \Phi^{\prime} 
\left( \Phi \Sigma_{W} \Phi^{\prime}  \right)^{-1} \Phi \bm{c}_{W,\bm{s}^{*}}.
\end{align*}
Since $C_{approx}$ underestimates the variance of $W(\bm{s})$ from $E[ \left( W(\bm{s}) - E[W(\bm{s})|\Phi \bm{W}] \right)^2 ] = C_{W}(\bm{s},\bm{s}) - \bm{c}_{W,\bm{s}}^{\prime} \Phi^{\prime} 
\left( \Phi \Sigma_{W} \Phi^{\prime}  \right)^{-1} \Phi \bm{c}_{W,\bm{s}} \ge 0$, Banerjee et al. (2013) defined
\begin{align}
C_{lp}(\bm{s},\bm{s}^{*}) = C_{approx}(\bm{s},\bm{s}^{*}) + \delta(\bm{s},\bm{s}^{*}) \left\{C_{W}(\bm{s},\bm{s}^{*}) - C_{approx}(\bm{s},\bm{s}^{*})\right\},
\label{modification1}
\end{align}
where $\delta(\bm{s},\bm{s}^{*})$ is 1 if $\bm{s} = \bm{s}^{*}$, otherwise 0. This modification is based on Finley et al. (2009). Let $\{W_{lp}(\bm{s})\}$ be a zero-mean Gaussian random field with the covariance function $C_{lp}$. $\Sigma_{approx}$, $\Sigma_{diag}$, and $\Sigma_{lp}$ denote the $n \times n$ covariance matrices with the $(i,j)$-th element of $C_{approx}(\bm{s}_i,\bm{s}_j)$, $\delta(\bm{s}_i,\bm{s}_j) \left(C_{W}(\bm{s}_i,\bm{s}_j) - C_{approx}(\bm{s}_i,\bm{s}_j) \right)$, and $C_{lp}(\bm{s}_i,\bm{s}_j)$ respectively. These matrix expressions are given by 
\begin{align*}
\Sigma_{approx} &= \Sigma_{W} \Phi^{\prime} \left( \Phi \Sigma_{W} \Phi^{\prime}  \right)^{-1} \Phi \Sigma_{W}, \\
\Sigma_{diag} &= \left\{ \Sigma_{W} - \Sigma_{W} \Phi^{\prime} 
\left( \Phi \Sigma_{W} \Phi^{\prime}  \right)^{-1} \Phi \Sigma_{W} \right\}  \circ \bm{I},
\intertext{and}
\Sigma_{lp} &= \Sigma_{approx} + \Sigma_{diag},
\end{align*}
where the notation '$\circ$' refers to the Hadamard product. Now, we replace $W(\bm{s})$ in \eqref{model} with $W_{lp}(\bm{s})$. Consequently, the covariance matrix of $\bm{Y}$ changes from $\Sigma_W + \tau^2 \bm{I}$ to $\Sigma_{lp} + \tau^2 \bm{I} = \Sigma_{W} \Phi^{\prime} 
\left( \Phi \Sigma_{W} \Phi^{\prime}  \right)^{-1} \Phi \Sigma_{W} + \Sigma_{diag}  + \tau^2 \bm{I}$ and the inverse matrix and determinant of $\Sigma_W + \tau^2 \bm{I}$ in the Bayesian inference and prediction of Section 2 become those of $\Sigma_{lp} + \tau^2 \bm{I}$. 
Using Harville (1997), we obtain
\begin{align}
\left\{ \Sigma_{W} \Phi^{\prime} 
\left( \Phi \Sigma_{W} \Phi^{\prime}  \right)^{-1} \Phi \Sigma_{W} + \Sigma_{diag}  + \tau^2 \bm{I} \right\}^{-1} = \left( \Sigma_{diag}  + \tau^2 \bm{I} \right)^{-1} - \left( \Sigma_{diag}  + \tau^2 \bm{I} \right)^{-1}  \Sigma_{W}\Phi^{\prime} & \notag \\
\times \left\{ \Phi \Sigma_{W} \Phi^{\prime} + \Phi \Sigma_{W} \left( \Sigma_{diag}  + \tau^2 \bm{I} \right)^{-1} \Sigma_{W} \Phi^{\prime} \right\}^{-1} \Phi \Sigma_{W} \left( \Sigma_{diag}  + \tau^2 \bm{I} \right)^{-1} &.
\label{invlp}
\end{align}
Similarly, from Harville (1997), the determinant can be calculated using 
\begin{align}
\left| \Sigma_{W} \Phi^{\prime} \left( \Phi \Sigma_{W} \Phi^{\prime}  \right)^{-1} \Phi \Sigma_{W} + \Sigma_{diag} + \tau^2 \bm{I} \right| =& \left| \Sigma_{diag} + \tau^2 \bm{I} \right| \left| \Phi \Sigma_{W} \Phi^{\prime} \right|^{-1} \notag \\
 &\times \left| \Phi \Sigma_{W} \Phi^{\prime} + \Phi \Sigma_{W} \left( \Sigma_{diag}  + \tau^2 \bm{I} \right)^{-1} \Sigma_{W} \Phi^{\prime}  \right|.
\label{detlp} 
\end{align}
The right-hand sides of \eqref{invlp} and \eqref{detlp} include the inversion and determinant of the $n \times n$ diagonal matrix $\Sigma_{diag} + \tau^2 \bm{I}$ and the $m \times m$ matrices, so that it is faster to conduct the Bayesian inference and prediction. If $\Phi$ is an $m \times n$ submatrix of an $n \times n$ permutation matrix, we obtain a predictive process whose knots are an $m$-dimensional subset of $\{ \bm{s}_1, \ldots, \bm{s}_n \}$. Therefore, the linear projection is an extension of predictive process models. Additionally, the linear projection approach avoids the knot selection problem of the predictive process.

Next, we explain the selection of $\Phi$ using the stochastic matrix approximation technique in Banerjee et al. (2013). It follows from Schmidt's approximation theorem (Stewart 1993; page 563) that $U_{m}^{\prime} = \argmin_{\Phi} \| \Sigma_{W} - \Sigma_{W}\Phi^{\prime}(\Phi \Sigma_{W} \Phi^{\prime})^{-1} \Phi \Sigma_{W} \|_{F}$ for fixed $m$ where $\|\cdot\|_{F}$ denotes the Frobenius norm for matrices and $U_{m}$ is the $n \times m$ matrix whose $i$th column vector is the eigenvector corresponding to the $i$th eigenvalue of $\Sigma_{W}$ in descending order of magnitude ($i = 1, \ldots, n$). However, the derivation of eigenvalues and eigenvectors of $\Sigma_{W}$ involves  $O(n^3)$ computations (Golub and Van Loan 1996). From $U_m U_m^{\prime} \Sigma_{W} = \Sigma_{W} U_m (U_m^{\prime} \Sigma_{W} U_m)^{-1} U_m^{\prime} \Sigma_{W}$, Banerjee et al. (2013) proposed the following algorithm to find $\Phi$ by diminishing $\| \Sigma_{W} - \Phi'\Phi \Sigma_{W} \|_F$ for any target error level on the basis of the appropriate modification of Algorithm 4.2 of Halko et al. (2011).
\par
\bigskip
\noindent
\textbf{Algorithm (Banerjee et al. 2013).} Given a target error $\epsilon > 0$ and $r \in \mathbb{N}$, find the $m \times n$ matrix $\Phi$ that satisfies $\| \Sigma_{W} - \Phi'\Phi \Sigma_{W} \|_F < \epsilon$ with probability $1- n/10^r$.

\noindent
\textit{Step} 1. Initialize $j = 0$ and $\Phi^{(0)}=[ \; ]$ (the $0 \times n$ empty matrix).

\noindent
\textit{Step} 2. Draw $r$ length-$n$ random vectors $\bm{\omega}^{(1)},\ldots,\bm{\omega}^{(r)}$ with independent entries from ${\cal N}(0,1)$.

\noindent
\textit{Step} 3. Compute $\bm{\kappa}^{(i)} = \Sigma_{W} \bm{\omega}^{(i)}$ for $i = 1,\ldots,r$.

\noindent
\textit{Step} 4. Check if $\max_{i = 1,\ldots,r}(\| \bm{\kappa}^{(i+j)} \|) < \{ (\pi/2)^{1/2} \epsilon \}/10$. If it holds, go to Step 11. Otherwise go to Step 5.

\noindent
\textit{Step} 5. Recompute j = j + 1, $\bm{\kappa}^{(j)} = \left[ \bm{I} - \{ \Phi^{(j-1)} \}^{\prime}  \Phi^{(j-1)} \right] \bm{\kappa}^{(j)}$, and $\bm{\phi}^{(j)} = \bm{\kappa}^{(j)}/\| \bm{\kappa}^{(j)} \|$.

\noindent
\textit{Step} 6. Set $\Phi^{(j)} = \left[ \{ \Phi^{(j-1)} \}^{\prime} \;\;\; \bm{\phi}^{(j)} \right]^{\prime}$.

\noindent
\textit{Step} 7. Draw a length-$n$ random vector $\bm{\omega}^{(j + r)}$ with independent entries from ${\cal N}(0,1)$.

\noindent
\textit{Step} 8. Compute $\bm{\kappa}^{(j+r)} = \left[ \bm{I} - \{ \Phi^{(j)} \}^{\prime}  \Phi^{(j)} \right] \Sigma_{W} \bm{\omega}^{(j + r)}$.

\noindent
\textit{Step} 9. Recompute $\bm{\kappa}^{(i)} = \bm{\kappa}^{(i)} - \bm{\phi}^{(j)} \{(\bm{\phi}^{(j)})^{\prime} \bm{\kappa}^{(i)} \}$ for $i = (j+1), \ldots, (j+r-1)$.

\noindent
\textit{Step} 10. Go back to the target error check in Step 4.

\noindent
\textit{Step} 11. If $j = 0$, output $\Phi = \left\{ \bm{\kappa}^{(1)}/\| \bm{\kappa}^{(1)} \| \right\}^{\prime}$; else output $\Phi = \Phi^{(j)}$.

Here, $\| \cdot \|$ denotes the Euclidean norm. Step 5 is not essential, but it ensures better stability when $\bm{\kappa}^{(j)}$ becomes very small (see Halko et al. 2011). Step 6 is the concatenation of the matrix and the vector. Banerjee et al. (2013) evaluated the linear projection approach using simulations and empirical examples and demonstrated that it achieved the better performance efficiently than the predictive process of Banerjee et al. (2008). 

\section{Modified linear projection}

As previously mentioned, the linear projection approach is not related to the knot selection problem unlike the predictive process. However, similarly to the predictive process, the linear projection approach is inaccurate when approximating local or small-scale dependences of the true covariance function $C_W$. In contrast, it is effective for the predictive process and the linear projection to capture large-scale spatial variations of $C_W$. 

\begin{figure}[h]
\begin{center}
 \subfigure[Linear Projection]{
  \includegraphics[width = 5.1cm]{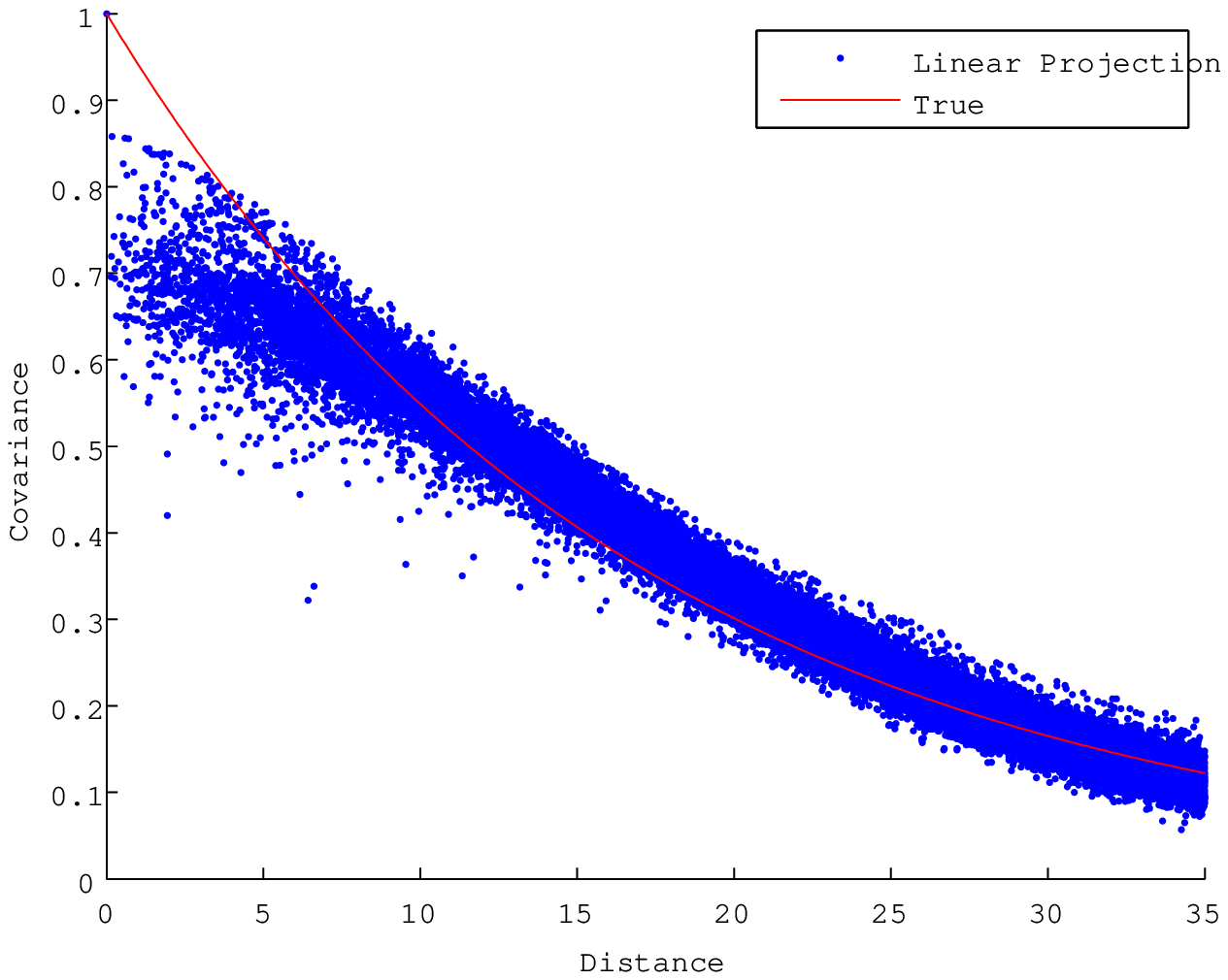}}
 \subfigure[Covariance Tapering]{ 
  \includegraphics[width = 5.1cm]{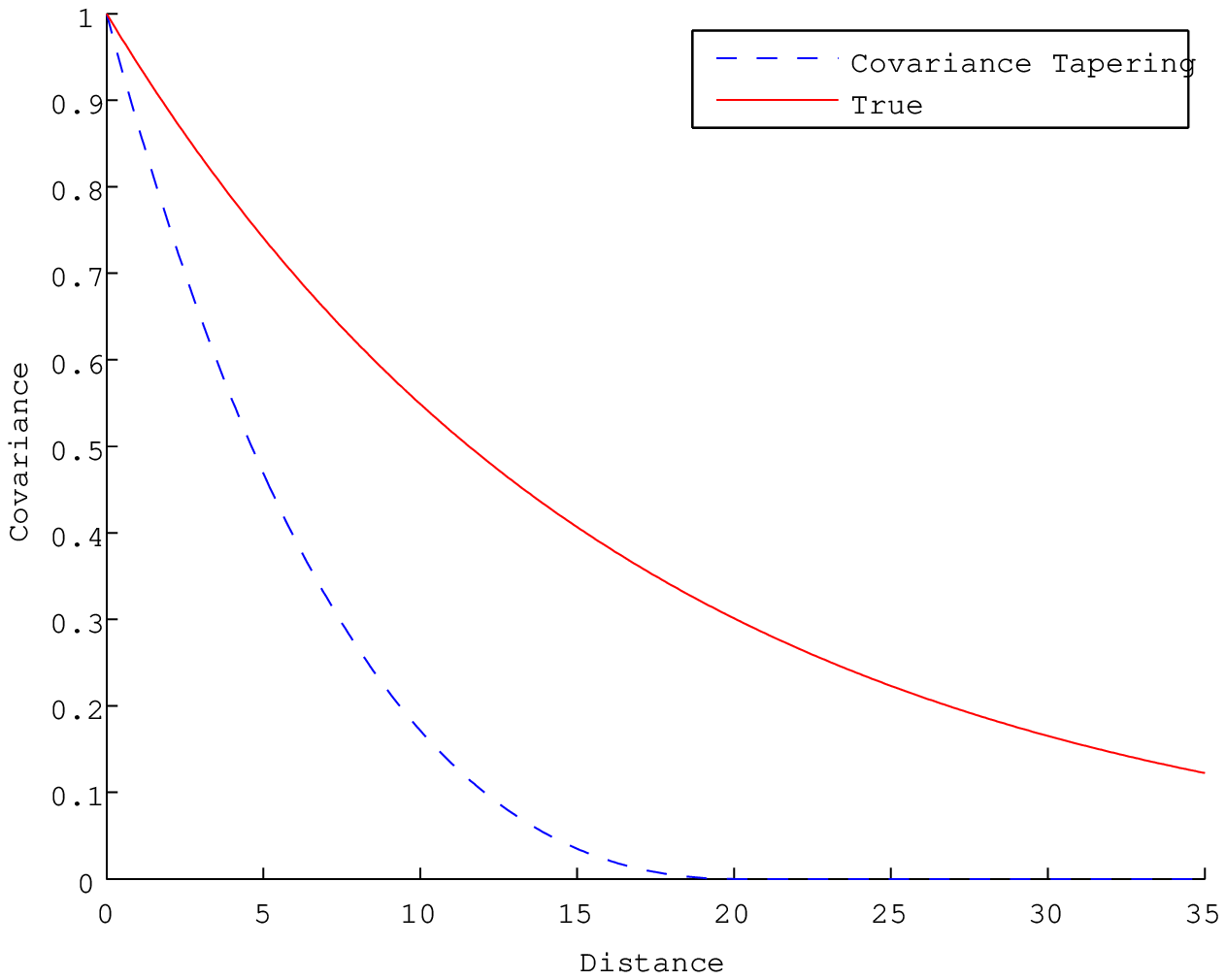}}
 \subfigure[Modified Linear Projection]{
  \includegraphics[width = 5.1cm]{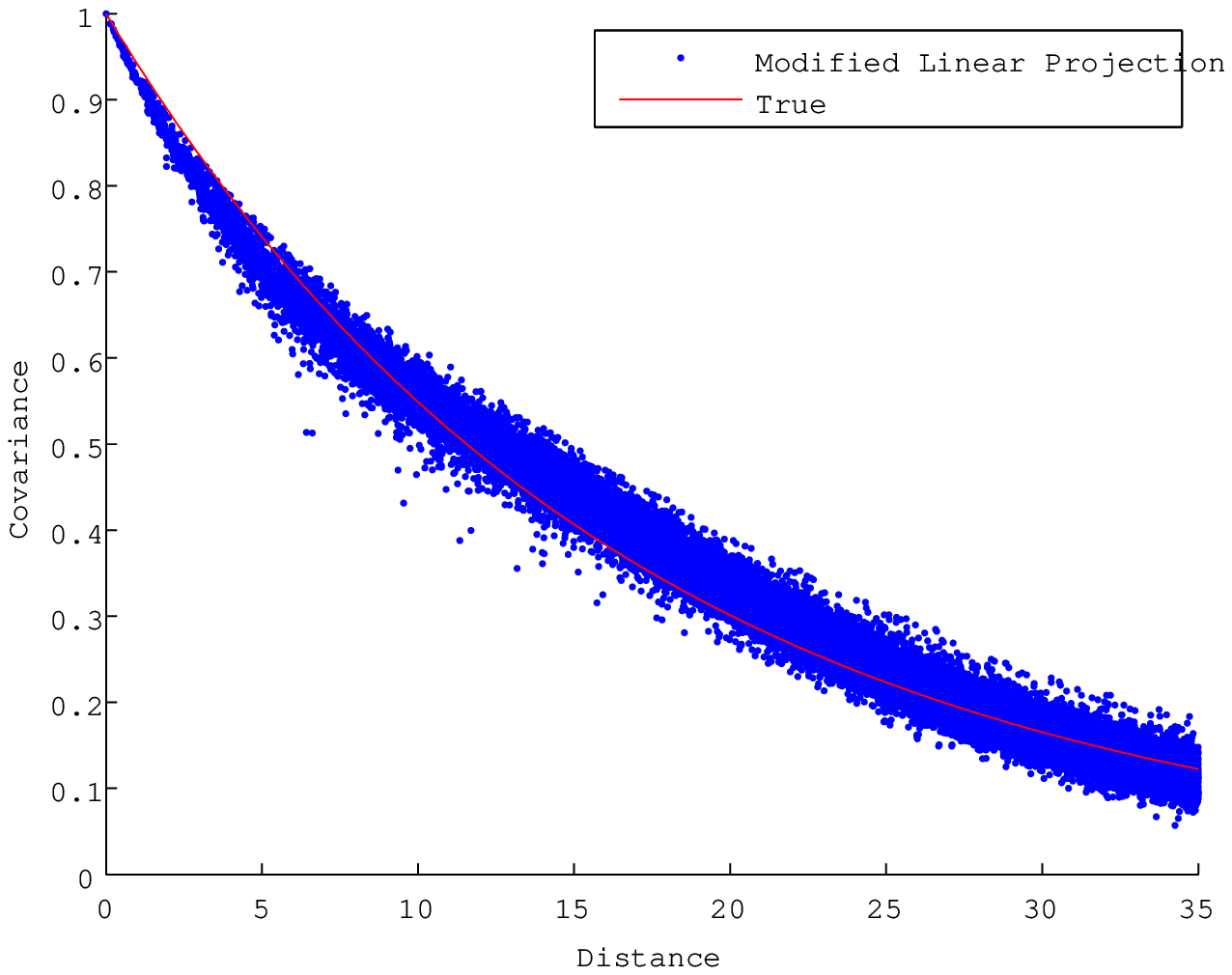}}
  \caption{Exponential covariance function $C_{W}(\bm{s}, \bm{s}^{*}) = \exp(-0.06\| \bm{s} - \bm{s}^{*} \|)$ (solid line) and three approximations. The true covariance matrix is generated using 500 random locations in $[0,100]\times[0,100]$. (a) Linear projection approach with $\epsilon = 200$ and $r=4$ (points). $m = 26$ was selected. (b) Covariance tapering using the spherical covariance function with $\gamma = 20$ (dotted line). (c) Modified linear projection using the linear projection with $\epsilon = 200$ and $r=4$ and the spherical covariance function with $\gamma = 20$ (points).} 
\end{center}
\end{figure}

Figure 1(a) shows a typical example to demonstrate that the improvement such as \eqref{modification1} is insufficient for modifying the approximation of small-scale dependence in the original covariance function. Using the linear projection, we obtain similar results for other covariance functions through some simulations (e.g., Gaussian covariance function and the Cauchy family (Gneiting and Schlather 2004)). 
 
Our proposed approach is a modification of the linear projection approach by the covariance tapering, which is based on the idea of the full scale approximation in Sang and Huang (2012). Before introducing our new approach, we review the covariance tapering which generates the sparse matrix approximation from the compactly supported correlation function and achieves the computational efficiency for analyzing large spatial data sets.

Let $K_{\gamma}(x)$ ($x \ge 0$ and $\gamma > 0$) be a compactly supported correlation function with $K_{\gamma}(0) = 1$ and $K_{\gamma}(x) = 0$ for $x \ge \gamma$. $K_{\gamma}(x)$ is called the taper function with a taper range $\gamma$. Some compactly supported correlation functions have been developed (see, e.g., Wendland 1995). For example, there are the spherical covariance function
\begin{align*}
K_{\gamma}(x) = \left(1-\frac{x}{\gamma} \right)^2_{+} \left( 1 + \frac{x}{2\gamma} \right)
\end{align*}
and
\begin{align}
K_{\gamma}(x) = \left(1-\frac{x}{\gamma} \right)^6_{+} \left(1+6\frac{x}{\gamma}+\frac{35x^2}{3\gamma^2} \right).
\label{Wendland_2}
\end{align}
Now, consider the product of the original covariance function and the taper function, that is
\begin{align*}
C_{ct}(\bm{s},\bm{s}^{*}) = C_{W}(\bm{s},\bm{s}^{*}) K_{\gamma}(\| \bm{s} - \bm{s}^{*} \|).
\end{align*}
Let $\{W_{ct}(\bm{s}) \}$ be a zero-mean Gaussian random field with the covariance function $C_{ct}$ and replace $W(\bm{s})$ in \eqref{model} with $W_{ct}(\bm{s})$. Then, $\Sigma_W + \tau^2 \bm{I}$ in the Bayesian inference and prediction of Section 2 becomes $\Sigma_W \circ \Sigma_{taper}  + \tau^2 \bm{I}$ where $(\Sigma_{taper})_{ij} = K_{\gamma}(\| \bm{s}_i - \bm{s}_j \|)$ ($i,j = 1,\ldots,n$). The resulting matrix $\Sigma_W \circ \Sigma_{taper}  + \tau^2 \bm{I}$ has many zero elements and is called a sparse matrix, so that we can use sparse matrix algorithms to efficiently handle the inverse matrix and determinant. 

From the definition of the covariance tapering, small-scale spatial dependence is well approximated, but large-scale dependence may not be appropriately accounted for (see Figure 1(b)). 
We introduce a modified linear projection approach to the covariance function of the original spatial process. It allows for efficient computations when using large spatial data sets. Define
\begin{align*}
C_{mlp}(\bm{s},\bm{s}^{*}) = C_{approx}(\bm{s},\bm{s}^{*}) + K_{\gamma}(\| \bm{s} - \bm{s}^{*} \|) \left\{ C_{W}(\bm{s},\bm{s}^{*}) - C_{approx}(\bm{s},\bm{s}^{*}) \right\}.
\end{align*}
$C_{mlp}$ is introduced by replacing the indicator function $\delta$ in \eqref{modification1} with the compactly supported correlation function $K_{\gamma}$ to incorporate the small-scale spatial dependence. 
Now, let $\Sigma_{sparse}$ and $\Sigma_{mlp}$ be the $n \times n$ Gram matrices with respect to $\bm{s}_1, \ldots, \bm{s}_n$ for $K_{\gamma} \times \left( C_{W} - C_{approx} \right)$ and $C_{mlp}$ respectively. These matrix expressions are given by 
\begin{align*}
\Sigma_{sparse} = \left\{ \Sigma_{W} - \Sigma_{W} \Phi^{\prime} 
\left( \Phi \Sigma_{W} \Phi^{\prime}  \right)^{-1} \Phi \Sigma_{W} \right\}  \circ \Sigma_{taper} \quad
\text{and} \quad
\Sigma_{mlp} = \Sigma_{approx} + \Sigma_{sparse}.
\end{align*}
The following proposition states the associated theoretical properties, which are used as conditions required in the expansion of the inversion and the determinant.
\par
\bigskip
\noindent
\textbf{Proposition 1}
\par
\noindent
(a) $\Phi \Sigma_W \Phi^{\prime}$ is positive definite.
\par
\noindent
(b) $\Sigma_{sparse} + \tau^2 \bm{I}$ is positive definite.
\par
\noindent
(c) $\Sigma_{mlp} + \tau^2 \bm{I}$ is positive definite.
\par
\noindent
(d) $\Phi \Sigma_{W} \Phi^{\prime} + \Phi \Sigma_{W} \left( \Sigma_{sparse}  + \tau^2 \bm{I} \right)^{-1} \Sigma_{W} \Phi^{\prime}$ is positive definite.

In the proof of Proposition 1(c), we prove that $\Sigma_{mlp}$ is positive semidefinite. An additional assumption on $\{ W(\bm{s}) \}$ yields the positive definiteness of $\Sigma_{mlp}$, but its proof is omitted for brevity. Consider a zero-mean Gaussian random field $\{W_{mlp}(\bm{s}) \}$ with the covariance function $C_{mlp}$ and replace $W(\bm{s})$ in \eqref{model} with $W_{mlp}(\bm{s})$. Consequently, $\left( \Sigma_W + \tau^2 \bm{I} \right)^{-1}$ and $\left| \Sigma_W + \tau^2 \bm{I} \right|$ in the Bayesian inference and prediction become the inverse matrix and determinant of $\Sigma_{mlp} + \tau^2 \bm{I} = \Sigma_{W} \Phi^{\prime} 
\left( \Phi \Sigma_{W} \Phi^{\prime}  \right)^{-1} \Phi \Sigma_{W} + \Sigma_{sparse}  + \tau^2 \bm{I}$ respectively. $\Sigma_{mlp}$ includes the original covariance matrix $\Sigma_{W}$ because $\Sigma_W = \Sigma_{mlp}$ if $m = n$. Similarly to the linear projection case, from Harville (1997) and Proposition 1, we can obtain
\begin{align}
\left\{ \Sigma_{W} \Phi^{\prime} 
\left( \Phi \Sigma_{W} \Phi^{\prime}  \right)^{-1} \Phi \Sigma_{W} + \Sigma_{sparse}  + \tau^2 \bm{I} \right\}^{-1} = \left( \Sigma_{sparse}  + \tau^2 \bm{I} \right)^{-1} - \left( \Sigma_{sparse}  + \tau^2 \bm{I} \right)^{-1}  \Sigma_{W}\Phi^{\prime} & \notag \\
\times \left\{ \Phi \Sigma_{W} \Phi^{\prime} + \Phi \Sigma_{W} \left( \Sigma_{sparse}  + \tau^2 \bm{I} \right)^{-1} \Sigma_{W} \Phi^{\prime} \right\}^{-1} \Phi \Sigma_{W} \left( \Sigma_{sparse}  + \tau^2 \bm{I} \right)^{-1} &
\label{invlp2}
\end{align}
and 
\begin{align}
\left| \Sigma_{W} \Phi^{\prime} \left( \Phi \Sigma_{W} \Phi^{\prime}  \right)^{-1} \Phi \Sigma_{W} + \Sigma_{sparse} + \tau^2 \bm{I} \right| =& \left| \Sigma_{sparse} + \tau^2 \bm{I} \right| \left| \Phi \Sigma_{W} \Phi^{\prime} \right|^{-1} \notag \\
 &\times \left| \Phi \Sigma_{W} \Phi^{\prime} + \Phi \Sigma_{W} \left( \Sigma_{sparse}  + \tau^2 \bm{I} \right)^{-1} \Sigma_{W} \Phi^{\prime}  \right|.
\label{detlp2} 
\end{align}
Now, we can treat the inverse matrix and determinant more quickly because \eqref{invlp2} and \eqref{detlp2} include the $n \times n$ sparse matrix $\Sigma_{sparse} + \tau^2 \bm{I}$ and $m \times m$ matrices. 
Figure 1(c) describes the good fitting of the modified linear projection to the original covariance function because the new approach uses the linear projection to capture large-scale spatial variations and the covariance tapering to capture small-scale local variations that are unexplained by the linear projection. In fact, the following proposition shows that the modified linear projection is superior to the linear projection in a sense of the Frobenius norm.
\par
\bigskip
\noindent
\textbf{Proposition 2}
\par
\noindent
Suppose that $K_{\gamma_1}(x) \le K_{\gamma_2}(x)$ for fixed $x \ge 0$. Then,
\[
\| \Sigma_W - \Sigma_{approx} \|_F \ge \| \Sigma_W - \Sigma_{lp} \|_F \ge \| \Sigma_W - \Sigma_{mlp,\gamma_1} \|_F \ge \| \Sigma_W - \Sigma_{mlp,\gamma_2} \|_F, 
\]
where $\Sigma_{mlp,\gamma_i} = \Sigma_{approx} + \left( \Sigma_{W} - \Sigma_{approx} \right)  \circ \Sigma_{taper,\gamma_i}$ and $(\Sigma_{taper,\gamma_i})_{kl} = K_{\gamma_i}(\| \bm{s}_k - \bm{s}_l \|)$ for $k,l = 1,\ldots,n$ and $i = 1,2$.

From Proposition 2, the approximation of the modified linear projection is better than that of the linear projection with respect to the Frobenius norm. Furthermore, it follows that the condition of Proposition 2 is satisfied for the spherical covariance function and the taper function \eqref{Wendland_2} when $\gamma_1 \le \gamma_2$. Then, as the taper range $\gamma$ increases, the approximation accuracy of the modified linear projection increases in a sense of the Frobenius norm. However, there is a trade-off between the magnitude of the taper range and the computational burden.

Finally, we now show that the approximation accuracy for the original covariance matrix controls the error in the probability density function of $\bm{Y}$. The next proposition is a corollary of Theorem 2 in Banerjee et al. (2013).
\par
\bigskip
\noindent
\textbf{Proposition 3}
\par
\noindent
Suppose that $\Sigma_{*}(A) = \Sigma_{W} \Phi^{\prime} \left( \Phi \Sigma_{W} \Phi^{\prime}  \right)^{-1} \Phi \Sigma_{W} + \left\{ \Sigma_{W} - \Sigma_{W} \Phi^{\prime} 
\left( \Phi \Sigma_{W} \Phi^{\prime}  \right)^{-1} \Phi \Sigma_{W} \right\}  \circ A$ and $A$ is an $n \times n$ positive definite matrix. Let $f = {\cal N}(X\bm{\beta}, \Sigma_W + \tau^2 \bm{I})$ be the probability density function of $\bm{Y}$ under the original model and $f_{*} = {\cal N}(X\bm{\beta}, \Sigma_{*}(A) + \tau^2 \bm{I})$  denotes its linear projection-type approximation. If $\| \Sigma_W - \Sigma_{*}(A) \|_F \le \epsilon$ for sufficiently small $\epsilon > 0$, then
\[
d_{KL}(f,f_{*}) \le \frac{n}{2} \left\{ \frac{\epsilon}{\tau^2} - \log \left( 1 - \frac{\epsilon}{\tau^2} \right)  \right\}, 
\]
where $d_{KL}$ denotes the Kullback-Leibler divergence between probability density functions.

In other words, the Kullback-Leibler divergence is of the same order as the error in the approximation of the original covariance matrix in terms of the Frobenius norm. Since most of our derivation is a straightforward application of Banerjee et al. (2013) without a small gap, we omit the proof of Proposition 3. Note that $\Sigma_{*}(\bm{I}) = \Sigma_{lp}$ and $\Sigma_{*}(\Sigma_{taper}) = \Sigma_{mlp}$. From Proposition 3, the error between the original probability density function and that of the modified linear projection has the sharp bound compared to the linear projection because $\| \Sigma_W - \Sigma_{lp} \|_F \ge \| \Sigma_W - \Sigma_{mlp} \|_F$ in Proposition 2. 

\section{Illustrative examples using simulated data}

This section illustrates our proposed method using simulated data and examines the effect of our modification using the compactly supported correlation function by comparing it with the linear projection and the covariance tapering. All computations were carried out using MATLAB functions {\tt sparse}, {\tt symamd}, and {\tt chol} on \rm{Linux  powered 2.50GHz Xeon processor with 64 Gbytes RAM}. 
The convergence diagnostics and the posterior summarization for MCMC were implemented by the R package {\tt CODA} (Plummer et al. 2006). In our simulations and the empirical study, the taper function \eqref{Wendland_2} was used for the covariance tapering and modified linear projection.

First, we investigated the performance of the proposed method through a simple simulation. Let $D = [0, 100]^2$ be the sampling domain and 2000 locations were sampled from a uniform distribution over $D$. We randomly selected 1500 locations for the estimation of parameters and DIC, while the rest were used for the calculation of the MSPE. We employed the Mat\'ern correlation function
\begin{align*}
\rho_{W} (\bm{s},\bm{s}_{*}  ; \nu, \lambda ) = \frac{1}{2^{\nu - 1} \Gamma(\nu)} \left( \frac{2 \nu^{1/2} \| \bm{s} - \bm{s}_{*} \|}{\lambda} \right)^{\nu} J_{\nu} \left(  \frac{2 \nu^{1/2} \| \bm{s} - \bm{s}_{*} \|}{\lambda}  \right), \quad \nu > 0, \; \lambda > 0, 
\end{align*}
where $\Gamma(\cdot)$ is the gamma function and $J_{\nu}(\cdot)$ is the modified Bessel function of the second kind of order $\nu$ (see Stein 1999). The spatial range parameter $\lambda$ controls the decay in spatial correlation and the smoothness parameter $\nu$ can be interpreted as the degree of the smoothness of the random field. For example, if $\nu=0.5$, the Mat\'ern correlation function is
\begin{align}
\rho_W(\bm{s},\bm{s}_{*}  ; \lambda) = \exp \left(- \frac{\sqrt{2} \| \bm{s} - \bm{s}_{*} \| }{\lambda} \right).
\label{exponential}
\end{align}
This is called the exponential covariance function and is widely used in many applications. The data were simulated from the model \eqref{model} with $\bm{\beta} = \bm{0}$, the exponential covariance function with $\sigma^2 = 0.5$, $\lambda = \sqrt{2}/0.06$ and $\sqrt{2}/0.3$, and nugget variance $\tau^2 = 1$. $\sigma^2$ and $\tau^2$ were the targets of the estimation and we assumed that the other parameters were known. When pairs of observations are more than 50 unit distant from each other in $\lambda = \sqrt{2}/0.06$, they have negligible ($<0.05$) correlation. This distance is called the effective range of the random field and 50 unit represents the random field with the strong spatial correlation. Similarly, the effective range in $\lambda = \sqrt{2}/0.3$ is 10 unit and it has the weak spatial correlation. For prior distributions, we assumed that $a_1 = 1$, $b_1 = 0.1$, $a_2 = 0.8$, and $b_2 = 0.1$ in \eqref{prior}. We ran the second and third stages in the MCMC algorithm presented in Section 2 for 50000 iterations, discarding the first 500 samples as burn-in periods. We applied the linear projection with $\epsilon = 200$ and $r =4$, the covariance tapering with $\gamma = 2.8$ and $20$, and the modified linear projection with $\epsilon = 200$, $r =4$, and $\gamma = 20$ in $\lambda = \sqrt{2}/0.06$. In $\lambda = \sqrt{2}/0.3$, the linear projection with $\epsilon = 150$, $400$, and $r =4$, the covariance tapering with $\gamma = 2.8$ and $10$, and the modified linear projection with $\epsilon = 400$, $r =4$, and $\gamma = 2.8$ were considered. 

The inefficiency factor (IF) is defined as $1 + 2 \sum_{t = 1}^{\infty} \rho(t)$ where $\rho(t)$ is the sample autocorrelation at lag $t$ for the parameter of interest. This factor is used to measure how well the MCMC mixes (e.g., Chib 2001). The smaller the inefficiency factor becomes, the closer the MCMC sampling is to the uncorrelated one. The computational time of each approach is relative to the time taken in the full model, scaled to 1. These times include the calculation of MSPE, DIC, and $\Phi$ selected by the algorithm in Section 3. Additionally, we described the rank of $\Phi$ required in the algorithm and the sparsity of the matrix measured by the percentage of zero elements in the off-diagonal elements of $\Sigma_{taper}$.

\begin{table}[ht]
  \caption{Summary of results from the first simulation in $\lambda = \sqrt{2}/0.06$.}
    \begin{tabular}{llcccccc}
    \toprule
          &       & $\tau^2$ & $\sigma^2$ &       & MSPE  & DIC   & Relative time \\
    \midrule
    \multicolumn{1}{l}{True value} & \multicolumn{1}{l}{} & 1     & 0.5   &       & -     & -     & - \\
    \multicolumn{1}{l}{} & \multicolumn{1}{l}{} &       &       &       &       &       &  \\
    \multicolumn{1}{l}{Original model} & \multicolumn{1}{l}{Mean} & 0.992 & 0.432 &       & \multirow{4}[0]{*}{1.138} & \multirow{4}[0]{*}{4472} & \multirow{4}[0]{*}{1} \\
    \multicolumn{1}{l}{} & \multicolumn{1}{l}{Stdev} & 0.043 & 0.090  &       &       &       &  \\
    \multicolumn{1}{l}{} & \multicolumn{1}{l}{95\% interval} & [0.911, 1.079] & [0.282, 0.633] &       &       &       &  \\
    \multicolumn{1}{l}{} & \multicolumn{1}{l}{IF} & 7.537 & 21.753 &       &       &       &  \\
    \multicolumn{1}{l}{} & \multicolumn{1}{l}{} &       &       &       &       &       &  \\
    \multicolumn{1}{l}{MLP} & \multicolumn{1}{l}{Mean} & 0.985 & 0.458 &       & \multirow{4}[0]{*}{1.157} & \multirow{4}[0]{*}{4479} & \multirow{4}[0]{*}{0.38} \\
    \multicolumn{1}{l}{($\epsilon = 200, \gamma = 2.8$)} & \multicolumn{1}{l}{Stdev} & 0.047 & 0.111 &       &       &       &  \\
    \multicolumn{1}{l}{} & \multicolumn{1}{l}{95\% interval} & [0.895, 1.077] & [0.281, 0.708] &       &       &       &  \\
    \multicolumn{1}{l}{} & \multicolumn{1}{l}{IF} & 12.919 & 35.117 &       &       &       &  \\
    \multicolumn{1}{l}{} & \multicolumn{1}{l}{} &       &       &       &       &       &  \\
    \multicolumn{1}{l}{LP } & \multicolumn{1}{l}{Mean} & 0.987 & 0.451 &       & \multirow{4}[0]{*}{1.158} & \multirow{4}[0]{*}{4480} & \multirow{4}[0]{*}{0.36} \\
    \multicolumn{1}{l}{($\epsilon = 200$)} & \multicolumn{1}{l}{Stdev} & 0.047 & 0.108 &       &       &       &  \\
    \multicolumn{1}{l}{} & \multicolumn{1}{l}{95\% interval} & [0.896, 1.079] & [0.276, 0.698] &       &       &       &  \\
    \multicolumn{1}{l}{} & \multicolumn{1}{l}{IF} & 13.612 & 31.757 &       &       &       &  \\
    \multicolumn{1}{l}{} & \multicolumn{1}{l}{} &       &       &       &       &       &  \\
    \multicolumn{1}{l}{CT } & \multicolumn{1}{l}{Mean} & 0.769 & 0.596 &       & \multirow{4}[0]{*}{1.399} & \multirow{4}[0]{*}{4702} & \multirow{4}[0]{*}{0.22} \\
    \multicolumn{1}{l}{($\gamma=2.8$)} & \multicolumn{1}{l}{Stdev} & 0.10  & 0.106  &       &       &       &  \\
    \multicolumn{1}{l}{} & \multicolumn{1}{l}{95\% interval} & [0.586, 0.979] & [0.388, 0.802] &       &       &       &  \\
    \multicolumn{1}{l}{} & \multicolumn{1}{l}{IF} & 46.439 & 48.492 &       &       &       &  \\
    \multicolumn{1}{l}{} & \multicolumn{1}{l}{} &       &       &       &       &       &  \\
    \multicolumn{1}{l}{CT } & \multicolumn{1}{l}{Mean} & 0.971 & 0.342 &       & \multirow{4}[0]{*}{1.158} & \multirow{4}[0]{*}{4514} & \multirow{4}[0]{*}{2.57} \\
    \multicolumn{1}{l}{($\gamma=20$)} & \multicolumn{1}{l}{Stdev} & 0.041 & 0.050  &       &       &       &  \\
    \multicolumn{1}{l}{} & \multicolumn{1}{l}{95\% interval} & [0.894, 1.056] & [0.255, 0.449] &       &       &       &  \\
    \multicolumn{1}{l}{} & \multicolumn{1}{l}{IF} & 5.408 & 8.832 &       &       &       &  \\
    \bottomrule
    \end{tabular}
    \\ MLP: modified linear projection; LP: linear projection; CT: covariance tapering. The required rank was 84 when $\epsilon = 200$. The sparsity was 0.24\% when $\gamma = 2.8$ and 10.71\% when $\gamma = 20$.
  \label{lambda006}
\end{table}

\begin{table}[H]
  \caption{Summary of results from the first simulation in $\lambda = \sqrt{2}/0.3$.}
    \begin{tabular}{llcccccc}
    \toprule
          &       & $\tau^2$ & $\sigma^2$ &       & MSPE  & DIC   & Relative time \\
    \midrule
    \multicolumn{1}{l}{True value} & \multicolumn{1}{l}{} & 1     & 0.5   &       & -     & -     & - \\
    \multicolumn{1}{l}{} & \multicolumn{1}{l}{} &       &       &       &       &       &  \\
    \multicolumn{1}{l}{Original model} & \multicolumn{1}{l}{Mean} & 1.018 & 0.427 &       & \multirow{4}[0]{*}{1.340 } & \multirow{4}[0]{*}{4717} & \multirow{4}[0]{*}{1} \\
    \multicolumn{1}{l}{} & \multicolumn{1}{l}{Stdev} & 0.061 & 0.074  &       &       &       &  \\
    \multicolumn{1}{l}{} & \multicolumn{1}{l}{95\% interval} & [0.90, 1.142] & [0.293, 0.582] &       &       &       &  \\
    \multicolumn{1}{l}{} & \multicolumn{1}{l}{IF} & 14.042 & 19.599 &       &       &       &  \\
    \multicolumn{1}{l}{} & \multicolumn{1}{l}{} &       &       &       &       &       &  \\
    \multicolumn{1}{l}{MLP} & \multicolumn{1}{l}{Mean} & 1.037 & 0.416 &       & \multirow{4}[0]{*}{1.408} & \multirow{4}[0]{*}{4754} & \multirow{4}[0]{*}{0.35} \\
    \multicolumn{1}{l}{($\epsilon = 400, \gamma = 2.8$)} & \multicolumn{1}{l}{Stdev} & 0.095 & 0.116 &       &       &       &  \\
    \multicolumn{1}{l}{} & \multicolumn{1}{l}{95\% interval} & [0.836, 1.211] & [0.224, 0.678] &       &       &       &  \\
    \multicolumn{1}{l}{} & \multicolumn{1}{l}{IF} & 47.139 & 55.115 &       &       &       &  \\
    \multicolumn{1}{l}{} & \multicolumn{1}{l}{} &       &       &       &       &       &  \\
    \multicolumn{1}{l}{LP } & \multicolumn{1}{l}{Mean} & 1.119 & 0.319 &       & \multirow{4}[0]{*}{1.433} & \multirow{4}[0]{*}{4763} & \multirow{4}[0]{*}{0.33} \\
    \multicolumn{1}{l}{($\epsilon = 400$)} & \multicolumn{1}{l}{Stdev} & 0.085 & 0.094 &       &       &       &  \\
    \multicolumn{1}{l}{} & \multicolumn{1}{l}{95\% interval} & [0.944, 1.277] & [0.161, 0.531] &       &       &       &  \\
    \multicolumn{1}{l}{} & \multicolumn{1}{l}{IF} & 219.804 & 407.408 &       &       &       &  \\
    \multicolumn{1}{l}{} & \multicolumn{1}{l}{} &       &       &       &       &       &  \\
    \multicolumn{1}{l}{CT } & \multicolumn{1}{l}{Mean} & 0.763 & 0.665 &       & \multirow{4}[0]{*}{1.444} & \multirow{4}[0]{*}{4772} & \multirow{4}[0]{*}{0.21} \\
    \multicolumn{1}{l}{($\gamma=2.8$)} & \multicolumn{1}{l}{Stdev} & 0.119  & 0.125  &       &       &       &  \\
    \multicolumn{1}{l}{} & \multicolumn{1}{l}{95\% interval} & [0.541, 1.011] & [0.412, 0.911] &       &       &       &  \\
    \multicolumn{1}{l}{} & \multicolumn{1}{l}{IF} & 64.035 & 66.709 &       &       &       &  \\
    \multicolumn{1}{l}{} & \multicolumn{1}{l}{} &       &       &       &       &       &  \\
    \multicolumn{1}{l}{LP } & \multicolumn{1}{l}{Mean} & 1.086 & 0.345 &       & \multirow{4}[0]{*}{1.351} & \multirow{4}[0]{*}{4724} & \multirow{4}[0]{*}{1.2} \\
    \multicolumn{1}{l}{($\epsilon = 150$)} & \multicolumn{1}{l}{Stdev} & 0.056 & 0.058 &       &       &       &  \\
    \multicolumn{1}{l}{} & \multicolumn{1}{l}{95\% interval} & [0.981, 1.20] & [0.238, 0.466] &       &       &       &  \\
    \multicolumn{1}{l}{} & \multicolumn{1}{l}{IF} & 37.746 & 149.150  &       &       &       &  \\
    \multicolumn{1}{l}{} & \multicolumn{1}{l}{} &       &       &       &       &       &  \\
    \multicolumn{1}{l}{CT } & \multicolumn{1}{l}{Mean} & 0.928 & 0.499 &       & \multirow{4}[0]{*}{1.367} & \multirow{4}[0]{*}{4727} & \multirow{4}[0]{*}{0.79} \\
    \multicolumn{1}{l}{($\gamma=10$)} & \multicolumn{1}{l}{Stdev} & 0.066  & 0.075  &       &       &       &  \\
    \multicolumn{1}{l}{} & \multicolumn{1}{l}{95\% interval} & [0.80, 1.063] & [0.359, 0.651] &       &       &       &  \\
    \multicolumn{1}{l}{} & \multicolumn{1}{l}{IF} & 16.551 & 20.983 &       &       &       &  \\
    \bottomrule
    \end{tabular}%
    \\ MLP: modified linear projection; LP: linear projection; CT: covariance tapering. The required rank was 87 when $\epsilon = 400$ and 510 when $\epsilon = 150$. The sparsity was 0.24\% when $\gamma = 2.8$ and 2.94\% when $\gamma = 10$.
  \label{lambda03}%
\end{table}%

Tables~\ref{lambda006} and \ref{lambda03} display the Bayesian posterior sample means, standard deviations, and 95\% credible intervals of the model parameters for each approach. Each approximation method required more time as $\epsilon$ decreased or $\gamma$ increased, which often offset the computational efficiency.
As shown in Table~\ref{lambda006}, the linear projection worked very well in the random field with the strong correlation. However, in the random field with the weak correlation, Table~\ref{lambda03} shows that the estimation of $\sigma^2$ using the linear projection was insufficient and the original model was superior to the linear projection even from a perspective of the calculation time. This is because the linear projection places a particular emphasis on fitting to the large-scale dependence. Since the covariance tapering has the property opposite to the linear projection, its performance is good except in the case where the linear projection is effective. Unlike the linear projection and the covariance tapering, the modified linear projection proposed in this paper performed well  regardless of the magnitude of the spatial correlation. The modified linear projection with appropriate taper range $\gamma$ improved the linear projection by adding a bit of time.

In the strong correlation case of the first simulation, the Frobenius norm of the difference between the original covariance matrix and the approximated one by the linear projection is 6.311 when $\epsilon = 200$. For the modified linear projection with $\epsilon = 200$, it is 6.033 when $\gamma = 2.8$ and 5.248 when $\gamma = 10$. In the weak correlation case of the first simulation, the Frobenius norm of the error is 16.198 for the linear projection with $\epsilon = 400$. For the modified linear projection with $\epsilon = 400$, it is 15.078 when $\gamma = 2.8$ and 12.524 when $\gamma = 10$. This supports the result of Proposition 2 and shows that the decrease of the Frobenius norm by the modified linear projection becomes large and the modification by the covariance tapering is effective for the random field where the small-scale dependence is dominant.
 
In the second simulation, we considered a nonstationary random field using the covariance function developed by Paciorek and Schervish (2006). The covariance function is
\begin{align}
C_W(\bm{s},\bm{s}_{*}) = \sigma^2 \frac{1}{2^{\nu -1} \Gamma(\nu)} | \Sigma_{D(\bm{s})} |^{\frac{1}{4}} 
| \Sigma_{D(\bm{s}_{*})} |^{\frac{1}{4}}  \left|   \frac{\Sigma_{D(\bm{s})} + \Sigma_{D(\bm{s}_{*})} }{2}  \right|^{- \frac{1}{2}} \left\{ 2 \sqrt{\nu d( \bm{s},\bm{s}_{*} )} \right\}^{\nu} J_{\nu} \left\{ 2 \sqrt{\nu d( \bm{s},\bm{s}_{*} )} \right\},
\label{nonstationary}
\end{align}
where $\Sigma_{D(\bm{s})}$ is a $d \times d$ positive definite matrix,
\begin{align*}
d( \bm{s},\bm{s}_{*} ) = ( \bm{s} - \bm{s}_{*} )^{\prime} \left( \frac{\Sigma_{D(\bm{s})} + \Sigma_{D(\bm{s}_{*})} }{2} \right)^{-1} ( \bm{s} - \bm{s}_{*} ),
\end{align*}
and $D(\bm{s})$ indicates the subregion which $\bm{s}$ belongs to. In the second simulation, we considered the sampling domain $D = [0, 500]^2$ and partitioned the entire region $D$ into two subregions $D_1 = [0, 250] \times [0, 500]$ and $D_2 = (250, 500] \times [0,500]$. 1000 locations were sampled from a uniform distribution over $D_i$ and split into 750 training sets and 250 test sets for $i = 1,2$. As a result, we obtained 1500 locations for training and 500 locations for testing. 
For the nonstationary covariance function \eqref{nonstationary}, the smoothness parameter $\nu$ was fixed to be 0.5 and we set $\Sigma_{D(\bm{s})} = \lambda_{D(\bm{s})}^2 \bm{I}$ where $D(\bm{s})$ is 1 if $\bm{s} \in D_1$, and 2 otherwise. This introduced the nonstationary random field that combined the stationary random field of the range parameter $\lambda_1$ over $D_1$ with that of the range parameter $\lambda_2$ over $D_2$. In Section 3 of Paciorek and Schervish (2006), there is an example of spatial data sets with different stationary covariance structures in the eastern and western regions similar to the nonstationary random field in our simulation. The data sets were simulated using the spatial linear regression model \eqref{model} with $x_1(\bm{s})=1$, $x_2(\bm{s})$ generated from the standard normal distribution, $\bm{\beta} = (1, 2)^{\prime}$, $\sigma^2 = 0.67$, $\tau^2 = 0.11$, $\lambda_1 = 1/0.08$, and $\lambda_2 = 1/0.3$. For prior distributions, we assumed that $\bm{\mu}_{\bm{\beta}} = (0.959, 1.972)^{\prime}$, $\Sigma_{\bm{\beta}} = 1000 \bm{I}$, $a_1 = 11$, $b_1 = 1.261$, $a_2 = 11$, and $b_2 = 6.305$. These hyperparameter choices were guided by the least squares estimator and an appropriate partition of the sample variance of its residual based on typical values in past empirical studies. 

In order to bypass the computational burden of the selection of $\Phi$ at each iteration, a discrete uniform distribution with atoms $\{ c_1, \ldots, c_{t_j} \}$ was taken as the prior distribution of $\lambda_j$ ($j = 1,2$) because we can precompute $\Phi$ using the algorithm presented in Section 3 for the correlation matrix of $W(\bm{s})$ with each distinct value of $\lambda_1 \in \{ c_1, \ldots, c_{t_1} \}$ and $\lambda_2 \in \{ c_1, \ldots, c_{t_2} \}$ prior to implementing the MCMC procedure. This strategy was proposed in Section 4 of Banerjee et al. (2013). In addition, Wikle (2010) used the discrete uniform distribution as the prior distribution for the range parameter. 

\noindent
\textit{Generation of $\lambda_1$.}

For $i = 1,\ldots, t_1$, the conditional posterior distribution of $\lambda_1$ is
\begin{align*}
&P(\lambda_1 = c_i | \bm{\beta}, \tau^2, \sigma^2,\lambda_2, \bm{Y}) \\
&= K_1  \left| \Sigma_W(\lambda_1 = c_i) + {\tau^2} \bm{I} \right|^{- \frac{1}{2}} \exp \left\{  - \frac{1}{2} (\bm{Y} - X \bm{\beta})^{\prime} \left( \Sigma_W(\lambda_1 = c_i) + {\tau^2} \bm{I} \right)^{-1} (\bm{Y} - X \bm{\beta})  \right\},
\end{align*}
where $\Sigma_W(\lambda_1 = c_i)$ denotes $\Sigma_W$ with $\lambda_1 = c_i$ and\\
 $K_1 = 1 \big/ \sum_{i = 1}^{t_1} \left| \Sigma_W(\lambda_1 = c_i) + {\tau^2} \bm{I} \right|^{- \frac{1}{2}}\exp \left\{  - \frac{1}{2} (\bm{Y} - X \bm{\beta})^{\prime} \left( \Sigma_W(\lambda_1 = c_i) + {\tau^2} \bm{I} \right)^{-1} (\bm{Y} - X \bm{\beta})  \right\}$.

\par
\noindent
\textit{Generation of $\lambda_2$.}

Similarly, for $i = 1,\ldots, t_2$, the conditional posterior distribution of $\lambda_2$ is
\begin{align*}
&P(\lambda_2 = c_i | \bm{\beta}, \tau^2, \sigma^2, \lambda_1, \bm{Y}) \\
&= K_2  \left| \Sigma_W(\lambda_2 = c_i) + {\tau^2} \bm{I} \right|^{- \frac{1}{2}} \exp \left\{  - \frac{1}{2} (\bm{Y} - X \bm{\beta})^{\prime} \left( \Sigma_W(\lambda_2 = c_i) + {\tau^2} \bm{I} \right)^{-1} (\bm{Y} - X \bm{\beta})  \right\},
\end{align*}
where $\Sigma_W(\lambda_2 = c_i)$ denotes $\Sigma_W$ with $\lambda_2 = c_i$ and\\
 $K_2 = 1 \big/ \sum_{i = 1}^{t_2} \left| \Sigma_W(\lambda_2 = c_i) + {\tau^2} \bm{I} \right|^{- \frac{1}{2}} \exp \left\{  - \frac{1}{2} (\bm{Y} - X \bm{\beta})^{\prime} \left( \Sigma_W(\lambda_2 = c_i) + {\tau^2} \bm{I} \right)^{-1} (\bm{Y} - X \bm{\beta})  \right\}$.

For $j = 1,2$, we set $c_i = 1/(0.02 i)$ ($i = 1,\ldots,t_j$) and $t_j = 25$ to choose a wide interval of the range parameters. The linear projection with $\epsilon = 350$ and $r =5$, the covariance tapering with $\gamma = 12$, and the modified linear projection with $\epsilon = 350$, $r =5$, and $\gamma = 12$ were applied. Using the MCMC algorithm described in Section 2, we sampled 8000 draws after the initial 300 samples were discarded as a burn-in period. 

\begin{table}[p]
  \caption{Summary of results from the second simulation.}
  \scalebox{0.95}{
    \begin{tabular}{rrrrrrr}
    \toprule
          & \multicolumn{1}{c}{True} & \multicolumn{1}{c}{} & \multicolumn{1}{c}{Original} & \multicolumn{1}{c}{MLP} & \multicolumn{1}{c}{LP} & \multicolumn{1}{c}{CT} \\
          & \multicolumn{1}{c}{value} &       &       & \multicolumn{1}{c}{($\epsilon = 350, \gamma = 12$)} & \multicolumn{1}{c}{($\epsilon = 350$)} & \multicolumn{1}{c}{($\gamma = 12$)} \\
          \midrule
    \multicolumn{1}{l}{} & \multicolumn{1}{l}{} & \multicolumn{1}{l}{} & \multicolumn{1}{c}{} & \multicolumn{1}{c}{} & \multicolumn{1}{c}{} & \multicolumn{1}{c}{} \\
    \multicolumn{1}{l}{\multirow{4}[0]{*}{$\beta_1$}} & \multicolumn{1}{l}{\multirow{4}[0]{*}{1}} & \multicolumn{1}{l}{Mean} & \multicolumn{1}{c}{0.949} & \multicolumn{1}{c}{0.952} & \multicolumn{1}{c}{0.933} & \multicolumn{1}{c}{0.962} \\
    \multicolumn{1}{l}{} & \multicolumn{1}{l}{} & \multicolumn{1}{l}{Stdev} & \multicolumn{1}{c}{0.029} & \multicolumn{1}{c}{0.030 } & \multicolumn{1}{c}{0.031 } & \multicolumn{1}{c}{0.023 } \\
    \multicolumn{1}{l}{} & \multicolumn{1}{l}{} & \multicolumn{1}{l}{95\% interval} & \multicolumn{1}{c}{[0.891, 1.004]} & \multicolumn{1}{c}{[0.892, 1.011]} & \multicolumn{1}{c}{[0.873, 0.994]} & \multicolumn{1}{c}{[0.916, 1.006]} \\
    \multicolumn{1}{l}{} & \multicolumn{1}{l}{} & \multicolumn{1}{l}{IF} & \multicolumn{1}{c}{1.425} & \multicolumn{1}{c}{5.641} & \multicolumn{1}{c}{2.530 } & \multicolumn{1}{c}{1.379 } \\
    \multicolumn{1}{l}{} & \multicolumn{1}{l}{} & \multicolumn{1}{l}{} & \multicolumn{1}{c}{} & \multicolumn{1}{c}{} & \multicolumn{1}{c}{} & \multicolumn{1}{c}{} \\
    \multicolumn{1}{l}{\multirow{4}[0]{*}{$\beta_2$}} & \multicolumn{1}{l}{\multirow{4}[0]{*}{2}} & \multicolumn{1}{l}{Mean} & \multicolumn{1}{c}{1.980 } & \multicolumn{1}{c}{1.972} & \multicolumn{1}{c}{1.970 } & \multicolumn{1}{c}{1.970 } \\
    \multicolumn{1}{l}{} & \multicolumn{1}{l}{} & \multicolumn{1}{l}{Stdev} & \multicolumn{1}{c}{0.020 } & \multicolumn{1}{c}{0.021} & \multicolumn{1}{c}{0.022} & \multicolumn{1}{c}{0.026} \\
    \multicolumn{1}{l}{} & \multicolumn{1}{l}{} & \multicolumn{1}{l}{95\% interval} & \multicolumn{1}{c}{[1.942, 2.019]} & \multicolumn{1}{c}{[1.930, 2.013]} & \multicolumn{1}{c}{[1.927, 2.012]} & \multicolumn{1}{c}{[1.929, 2.011]} \\
    \multicolumn{1}{l}{} & \multicolumn{1}{l}{} & \multicolumn{1}{l}{IF} & \multicolumn{1}{c}{1.0 } & \multicolumn{1}{c}{1.0 } & \multicolumn{1}{c}{1.0 } & \multicolumn{1}{c}{1.0 } \\
    \multicolumn{1}{l}{} & \multicolumn{1}{l}{} & \multicolumn{1}{l}{} & \multicolumn{1}{c}{} & \multicolumn{1}{c}{} & \multicolumn{1}{c}{} & \multicolumn{1}{c}{} \\
    \multicolumn{1}{l}{\multirow{4}[0]{*}{$\tau^2$}} & \multicolumn{1}{l}{\multirow{4}[0]{*}{0.11}} & \multicolumn{1}{l}{Mean} & \multicolumn{1}{c}{0.138} & \multicolumn{1}{c}{0.139} & \multicolumn{1}{c}{0.214} & \multicolumn{1}{c}{0.120 } \\
    \multicolumn{1}{l}{} & \multicolumn{1}{l}{} & \multicolumn{1}{l}{Stdev} & \multicolumn{1}{c}{0.031 } & \multicolumn{1}{c}{0.039 } & \multicolumn{1}{c}{0.062 } & \multicolumn{1}{c}{0.029 } \\
    \multicolumn{1}{l}{} & \multicolumn{1}{l}{} & \multicolumn{1}{l}{95\% interval} & \multicolumn{1}{c}{[0.086, 0.203]} & \multicolumn{1}{c}{[0.079, 0.223]} & \multicolumn{1}{c}{[0.099, 0.327]} & \multicolumn{1}{c}{[0.073, 0.185]} \\
    \multicolumn{1}{l}{} & \multicolumn{1}{l}{} & \multicolumn{1}{l}{IF} & \multicolumn{1}{c}{36.209} & \multicolumn{1}{c}{74.679} & \multicolumn{1}{c}{131.557} & \multicolumn{1}{c}{36.937} \\
    \multicolumn{1}{l}{} & \multicolumn{1}{l}{} & \multicolumn{1}{l}{} & \multicolumn{1}{c}{} & \multicolumn{1}{c}{} & \multicolumn{1}{c}{} & \multicolumn{1}{c}{} \\   
    \multicolumn{1}{l}{\multirow{4}[0]{*}{$\sigma^2$}} & \multicolumn{1}{l}{\multirow{4}[0]{*}{0.67}} & \multicolumn{1}{l}{Mean} & \multicolumn{1}{c}{0.628} & \multicolumn{1}{c}{0.643} & \multicolumn{1}{c}{0.589} & \multicolumn{1}{c}{0.629} \\
    \multicolumn{1}{l}{} & \multicolumn{1}{l}{} & \multicolumn{1}{l}{Stdev} & \multicolumn{1}{c}{0.041} & \multicolumn{1}{c}{0.049} & \multicolumn{1}{c}{0.080 } & \multicolumn{1}{c}{0.039 } \\
    \multicolumn{1}{l}{} & \multicolumn{1}{l}{} & \multicolumn{1}{l}{95\% interval} & \multicolumn{1}{c}{[0.546, 0.706]} & \multicolumn{1}{c}{[0.546, 0.738]} & \multicolumn{1}{c}{[0.443, 0.743]} & \multicolumn{1}{c}{[0.548, 0.707]} \\
    \multicolumn{1}{l}{} & \multicolumn{1}{l}{} & \multicolumn{1}{l}{IF} & \multicolumn{1}{c}{26.513} & \multicolumn{1}{c}{40.110 } & \multicolumn{1}{c}{129.658} & \multicolumn{1}{c}{26.337} \\
    \multicolumn{1}{l}{} & \multicolumn{1}{l}{} & \multicolumn{1}{l}{} & \multicolumn{1}{c}{} & \multicolumn{1}{c}{} & \multicolumn{1}{c}{} & \multicolumn{1}{c}{} \\
    \multicolumn{1}{l}{\multirow{4}[0]{*}{$\lambda_1$}} & \multicolumn{1}{l}{\multirow{4}[0]{*}{1/0.08}} & \multicolumn{1}{l}{Mean} & \multicolumn{1}{c}{13.944} & \multicolumn{1}{c}{18.257} & \multicolumn{1}{c}{24.368} & \multicolumn{1}{c}{36.215} \\
    \multicolumn{1}{l}{} & \multicolumn{1}{l}{} & \multicolumn{1}{l}{Stdev} & \multicolumn{1}{c}{2.075} & \multicolumn{1}{c}{3.521 } & \multicolumn{1}{c}{2.206 } & \multicolumn{1}{c}{15.197 } \\
    \multicolumn{1}{l}{} & \multicolumn{1}{l}{} & \multicolumn{1}{l}{95\% interval} & \multicolumn{1}{c}{[12.50, 16.667]} & \multicolumn{1}{c}{[12.50, 25.0]} & \multicolumn{1}{c}{[16.667, 25.0]} & \multicolumn{1}{c}{[10.0, 50.0]} \\
    \multicolumn{1}{l}{} & \multicolumn{1}{l}{} & \multicolumn{1}{l}{IF} & \multicolumn{1}{c}{14.361} & \multicolumn{1}{c}{43.472} & \multicolumn{1}{c}{37.155} & \multicolumn{1}{c}{2.208} \\
          & \multicolumn{1}{l}{} &       &       &       &       &  \\
    \multicolumn{1}{l}{\multirow{4}[0]{*}{$\lambda_2$}} & \multicolumn{1}{l}{\multirow{4}[0]{*}{1/0.3}} & \multicolumn{1}{l}{Mean} & \multicolumn{1}{c}{2.714} & \multicolumn{1}{c}{3.045} & \multicolumn{1}{c}{3.341} & \multicolumn{1}{c}{4.493} \\
    \multicolumn{1}{l}{} & \multicolumn{1}{l}{} & \multicolumn{1}{l}{Stdev} & \multicolumn{1}{c}{0.523} & \multicolumn{1}{c}{1.089} & \multicolumn{1}{c}{1.020 } & \multicolumn{1}{c}{2.075 } \\
    \multicolumn{1}{l}{} & \multicolumn{1}{l}{} & \multicolumn{1}{l}{95\% interval} & \multicolumn{1}{c}{[2.0, 3.846]} & \multicolumn{1}{c}{[2.174, 5.556]} & \multicolumn{1}{c}{[2.381, 5.556]} & \multicolumn{1}{c}{[2.083, 10.0]} \\
    \multicolumn{1}{l}{} & \multicolumn{1}{l}{} & \multicolumn{1}{l}{IF} & \multicolumn{1}{c}{1.131} & \multicolumn{1}{c}{1.269} & \multicolumn{1}{c}{2.952} & \multicolumn{1}{c}{1.263} \\
          & \multicolumn{1}{l}{} &       &       &       &       &  \\
    \multicolumn{1}{l}{MSPE} & \multicolumn{1}{c}{-} &       & \multicolumn{1}{c}{0.664} & \multicolumn{1}{c}{0.681} & \multicolumn{1}{c}{0.710 } & \multicolumn{1}{c}{0.708 } \\
    \multicolumn{1}{l}{DIC} & \multicolumn{1}{c}{-} &       & \multicolumn{1}{c}{3701} & \multicolumn{1}{c}{3724} & \multicolumn{1}{c}{3746} & \multicolumn{1}{c}{3780} \\
    \multicolumn{1}{l}{Relative time} & \multicolumn{1}{c}{-} &       & \multicolumn{1}{c}{1} & \multicolumn{1}{c}{0.54} & \multicolumn{1}{c}{0.52} & \multicolumn{1}{c}{0.36} \\
    \bottomrule
    \end{tabular}%
    }
    \\ MLP: modified linear projection; LP: linear projection; CT: covariance tapering. Average required rank and its 95\% interval were 23.75 and [1, 94.50] when $\epsilon = 350$. The sparsity was 0.17\% when $\gamma = 12$.
  \label{nonst}%
\end{table}%

The results of the simulation are summarized in Table~\ref{nonst}. For $\sigma^2$ and $\tau^2$, the linear projection shows high inefficiency factors and has discrepancies from the posterior means of the original model. This implies that 
the modification by the indicator function is not sufficient for small-scale variations. For $\beta_1$ and $\lambda_1$, the 95\% interval of the linear projection does not include the true value. The covariance tapering is highly computationally efficient, but the estimations of the range parameters are inaccurate. However, our proposed method modifies these drawbacks and works well compared to the linear projection and the covariance tapering. We obtained similar results with other settings, but these are not reported here. 

\section{Empirical study}

\begin{figure}[h]
\begin{center}
  \scalebox{0.6}{\includegraphics{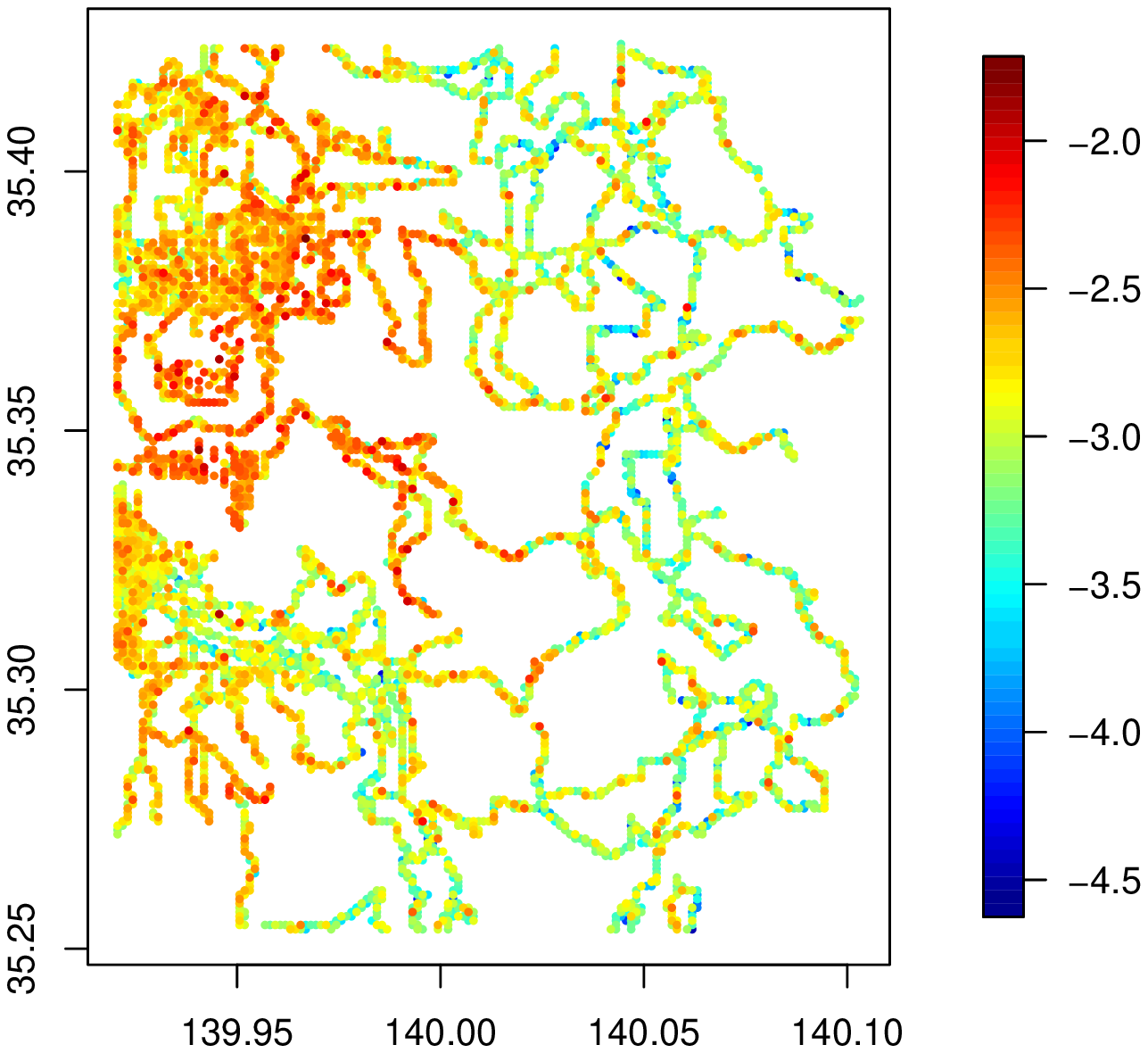}}
  \caption{The logarithmic transformation of air dose rates at 5557 sampling points in Chiba prefecture.}
  \label{originaldata}%
\end{center}
\end{figure}

In this section, we discuss the results when we applied our proposed modified linear projection method to air dose rates in Chiba prefecture. The data are created based on the results of the vehicle-borne survey conducted by the Japanese Ministry of Education, Culture, Sports, Science and Technology from November 5 to December 10, 2012. This data set consists of air dose rates (microsievert per hour) with longitudes, latitudes, and distances from the Fukushima Dai-ichi Nuclear Power Plant (NPP) (km) at 47470 sampling points and is obtained from the Environment Monitoring Database for the Distribution of Radioactive Substances Released by the TEPCO Fukushima Dai-ichi NPP Accident at \textit{http://radb.jaea.go.jp/mapdb/en/}. These are spatio-temporal data because they were observed on irregularly spaced locations at discrete time points. However, we have considered the data set to be spatial by assuming that the air dose rate trend does not fluctuate largely over a short period. To assume the Gaussian process over the whole region, we selected 5557 points inside the rectangular region $[139.920625, 140.103125] \times[35.25375, 35.424584]$. Figure \ref{originaldata} shows the logarithmic transformation of these spatial data sets. To understand the entire trend for the scattered observations, we attempted to make the prediction surface using representative points of the predictive distribution. 

We split 5557 data points into a training set of 5000 observations and a test set of 557 observations. To account for the mean component and maintain normality, we considered the spatial regression model \eqref{model} with the logarithmic transformation $Y(\bm{s})$ of the air dose rate, $x_1(\bm{s}) = 1$, and the distance from the Fukushima Dai-ichi NPP $x_2(\bm{s})$. 

The MCMC algorithm was similar to the second simulation presented in Section 5. The hyperparameters of the prior distributions for $\bm{\beta}=(\beta_1, \beta_2)^{\prime}$, $\sigma^2$, and $\tau^2$ were assumed to be $\bm{\mu}_{\bm{\beta}} = (-2.5849, -0.0012)^{\prime}$, $\Sigma_{\bm{\beta}} = 1000 \bm{I}$, $a_1 = 11$, $b_1 = 0.2121$, $a_2 = 11$, and $b_2 = 1.0607$ in the same way as the second simulation. Moreover, we conducted some trial runs using a training subset. The results led us to use the exponential covariance function \eqref{exponential} and the discrete uniform distribution with $c_i = 1/(0.01i)$ ($i = 1,\ldots,60$) for the prior distribution of $\lambda$. To implement the modified linear projection, we conducted a pilot analysis using training and test subsets for various choices of $\epsilon$ and $\gamma$. Weighing the trade-off between the prediction accuracy and run time, we determined that $\epsilon = 1200$, $r = 5$, and $\gamma = 0.5$ were appropriate choices. However, for $\epsilon = 1200$, the linear projection caused poor mixing of the sampling of $\lambda$ because of the weak spatial correlation of the logarithmic transformation of the air dose rate. Hence, $\epsilon = 300$ and $r=5$ were selected for the linear projection, which resulted in a longer run time than that of the modified linear projection. 

To run the MCMC algorithm, we drew 4100 samples and discarded 100 samples as a burn-in period. The predictive surfaces were generated by considering the predictive distribution at $31 \times 31$ prediction points, which overlaid the  sampling domain, and calculating the mean, 5\%, and 95\% quantiles of samples from the predictive distribution. The mean of the predictive distribution was used as the single point predictor and the 5\% and 95\% quantiles of the predictive distribution served as measures of uncertainty of the mean of the predictive distribution. Note that the calculation time for each method includes the generation of the predictive surfaces as well as the Bayesian estimation and prediction. 

\begin{table}[H]
  \caption{The result of the Bayesian analysis in the empirical data example.}
    \scalebox{0.95}{
    \begin{tabular}{rrcccc}
    \toprule
          & \multicolumn{1}{c}{} & Stationary model & MLP   & LP    & CT \\
          &       &  & ($\epsilon = 1200, \gamma = 0.5$) & ($\epsilon = 300$) & ($\gamma = 0.5$) \\
          \midrule
    \multicolumn{1}{l}{} & \multicolumn{1}{l}{} &       &       &       &  \\
    \multicolumn{1}{l}{\multirow{4}[0]{*}{$\beta_1$}} & \multicolumn{1}{l}{Mean} & -3.325 & -1.559 & -3.295 & -2.971 \\
    \multicolumn{1}{l}{} & \multicolumn{1}{l}{Stdev} & 2.986 & 2.897  & 1.441  & 0.380  \\
    \multicolumn{1}{l}{} & \multicolumn{1}{l}{95\% interval} & [-9.206, 2.615] & [-7.248, 3.974] & [-6.118, -0.434] & [-3.696, -2.218] \\
    \multicolumn{1}{l}{} & \multicolumn{1}{l}{IF} & 0.970  & 1.137 & 1.0   & 0.867  \\
    \multicolumn{1}{l}{} & \multicolumn{1}{l}{} &       &       &       &  \\
    \multicolumn{1}{l}{\multirow{4}[0]{*}{$\beta_2$}} & \multicolumn{1}{l}{Mean} & 0.001  & -0.006 & 0.001  & 0.0003  \\
    \multicolumn{1}{l}{} & \multicolumn{1}{l}{Stdev} & 0.012  & 0.012 & 0.006 & 0.002 \\
    \multicolumn{1}{l}{} & \multicolumn{1}{l}{95\% interval} & [-0.023, 0.025] & [-0.028, 0.018] & [-0.010, 0.013] & [-0.003, 0.003] \\
    \multicolumn{1}{l}{} & \multicolumn{1}{l}{IF} & 0.974  & 1.137  & 1.0   & 0.868  \\
    \multicolumn{1}{l}{} & \multicolumn{1}{l}{} &       &       &       &  \\
     \multicolumn{1}{l}{\multirow{4}[0]{*}{$\tau^2$}} & \multicolumn{1}{l}{Mean} & 0.070  & 0.055 & 0.068 & 0.049  \\
    \multicolumn{1}{l}{} & \multicolumn{1}{l}{Stdev} & 0.002  & 0.002  & 0.002  & 0.002  \\
    \multicolumn{1}{l}{} & \multicolumn{1}{l}{95\% interval} & [0.066, 0.073] & [0.051, 0.058] & [0.065, 0.072] & [0.046, 0.052] \\
    \multicolumn{1}{l}{} & \multicolumn{1}{l}{IF} & 11.872 & 15.055 & 13.0  & 9.452 \\
    \multicolumn{1}{l}{} & \multicolumn{1}{l}{} &       &       &       &  \\
    \multicolumn{1}{l}{\multirow{4}[0]{*}{$\sigma^2$}} & \multicolumn{1}{l}{Mean} & 0.077 & 0.110  & 0.446 & 0.070  \\
    \multicolumn{1}{l}{} & \multicolumn{1}{l}{Stdev} & 0.018 & 0.009 & 0.005  & 0.003  \\
    \multicolumn{1}{l}{} & \multicolumn{1}{l}{95\% interval} & [0.050, 0.121] & [0.093, 0.129] & [0.036, 0.055] & [0.065, 0.076] \\
    \multicolumn{1}{l}{} & \multicolumn{1}{l}{IF} & 38.980  & 11.472  & 6.825 & 7.074 \\
    \multicolumn{1}{l}{} & \multicolumn{1}{l}{} &       &       &       &  \\
    \multicolumn{1}{l}{\multirow{4}[0]{*}{$\lambda$}} & \multicolumn{1}{l}{Mean} & 4.645 & 4.187 & 1.735 & 35.601 \\
    \multicolumn{1}{l}{} & \multicolumn{1}{l}{Stdev} & 1.594 & 0.308  & 0.182  & 31.961  \\
    \multicolumn{1}{l}{} & \multicolumn{1}{l}{95\% interval} & [2.381, 8.333] & [3.704, 5.556] & [1.667, 2.041] & [4.762, 100.0] \\
    \multicolumn{1}{l}{} & \multicolumn{1}{l}{IF} & 32.214 & 4.20  & 4.912 & 1.202 \\
          &       & \multicolumn{1}{r}{} & \multicolumn{1}{r}{} & \multicolumn{1}{r}{} & \multicolumn{1}{r}{} \\
    \multicolumn{1}{l}{MSPE} &       & 0.076 & 0.071 & 0.079  & 0.074  \\
    \multicolumn{1}{l}{DIC} &       & 1703  & 1598  & 1747  & 2551 \\
    \multicolumn{1}{l}{Relative time} &       & 1     & 0.51  & 0.59  & 0.34 \\
    \bottomrule
    \end{tabular}%
    }
     \\ MLP: modified linear projection; LP: linear projection; CT: covariance tapering. Average required rank and its 95\% interval were 45.133 and [8, 77] when $\epsilon = 1200$ and 187.88 and [26, 335.50] when $\epsilon = 300$. The sparsity was 0.46\% when $\gamma = 0.5$.
  \label{empirical}%
\end{table}%

\begin{figure}[H]
\begin{center}
  \scalebox{0.45}{\includegraphics{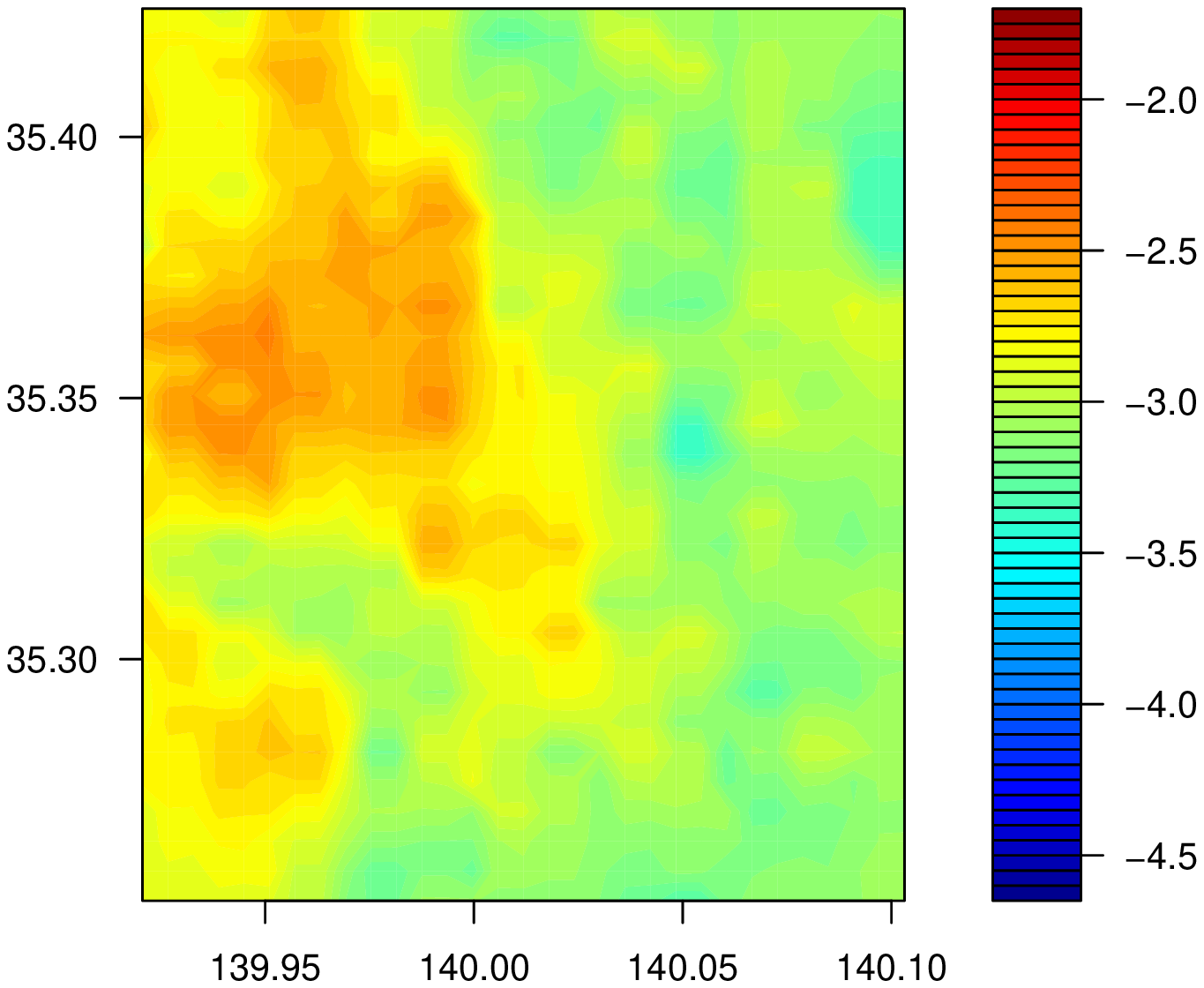}}
  \scalebox{0.45}{\includegraphics{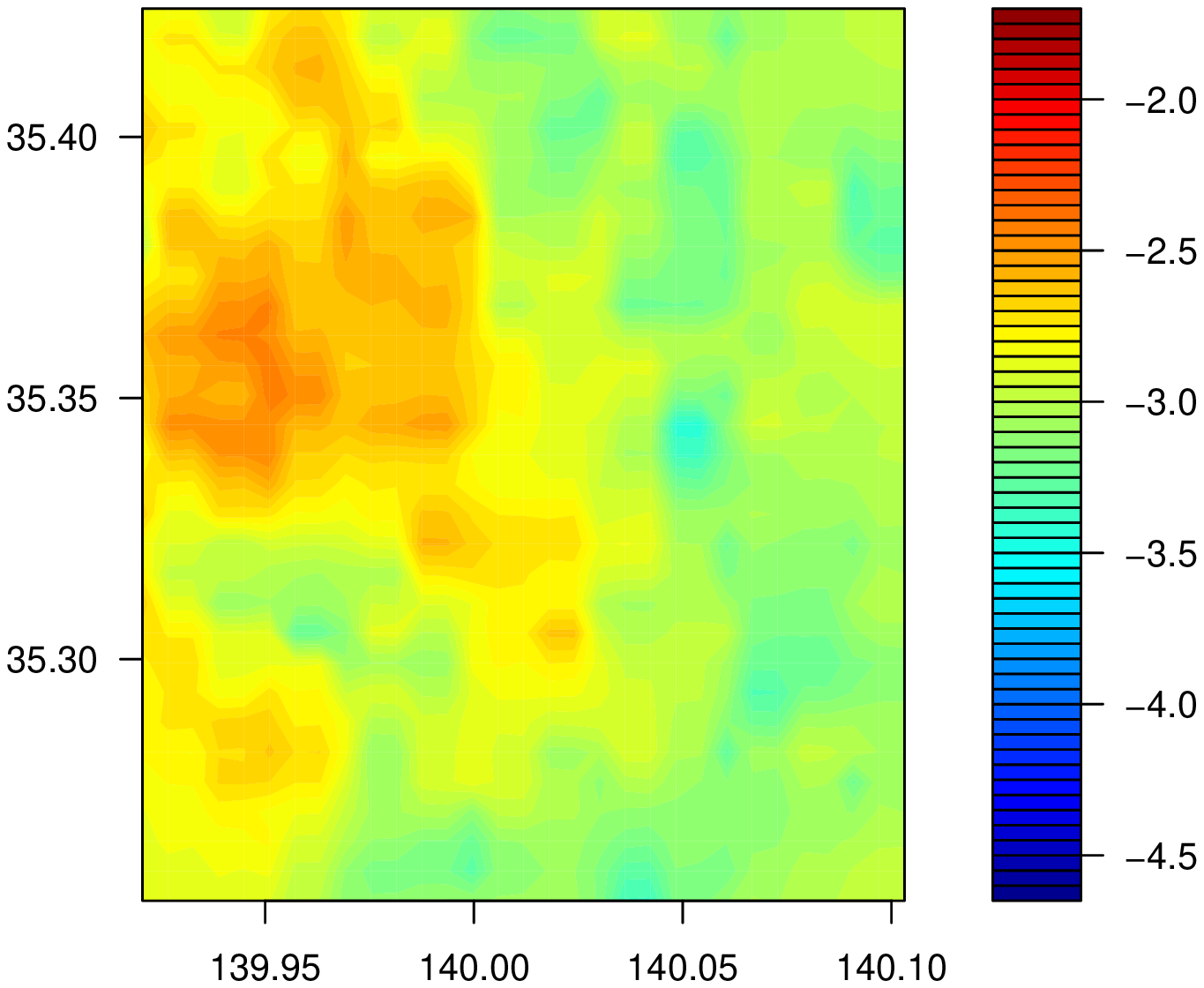}}
  \scalebox{0.45}{\includegraphics{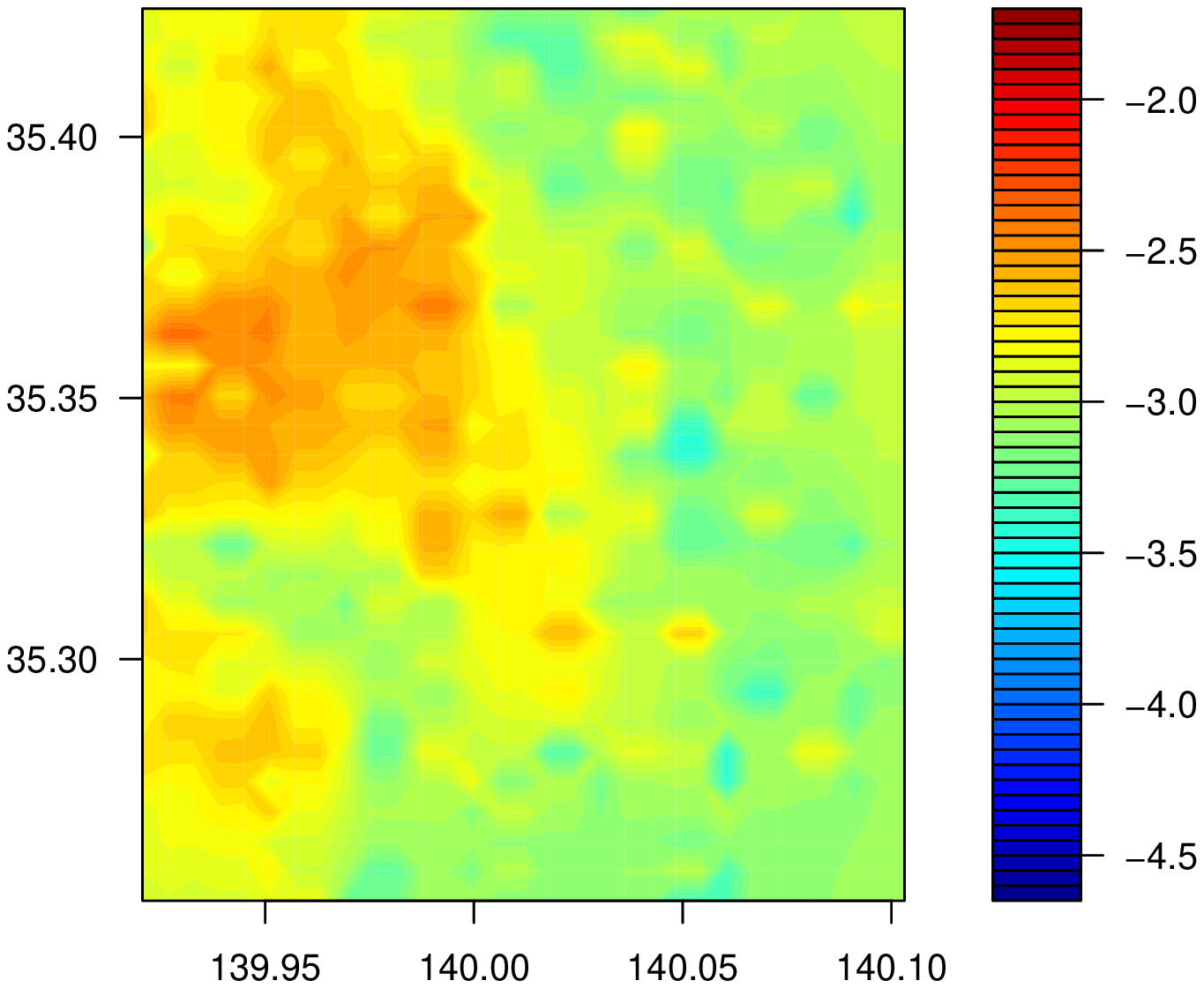}}
  \scalebox{0.45}{\includegraphics{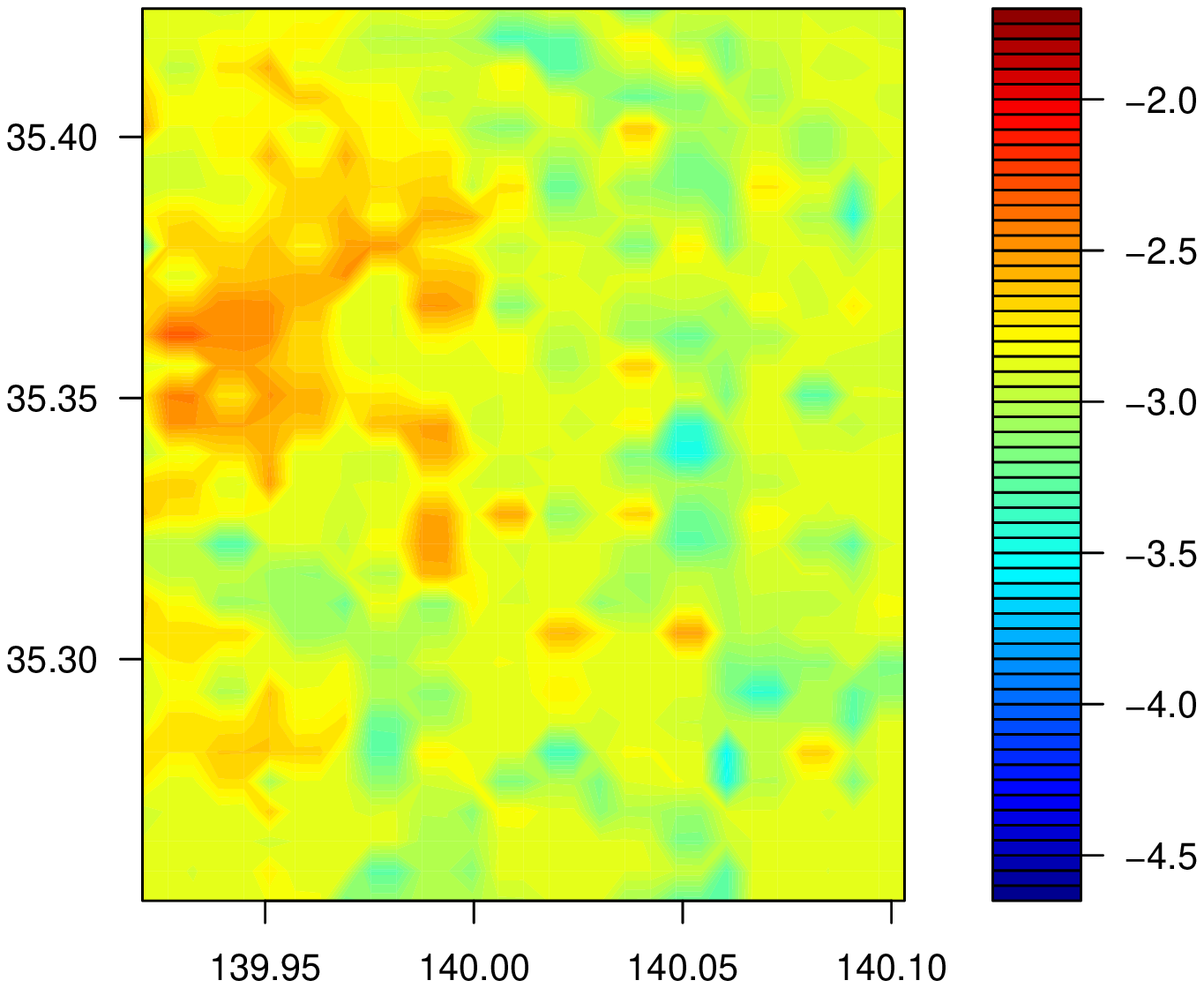}}
  \caption{Mean of the predictive distribution. We adopt the stationary model and the linear projection for the left and right column in the top row respectively. The modified linear projection and the covariance tapering are used for the left and right column in the bottom row respectively.}
    \label{meansurface}%
\end{center}
\end{figure}

\begin{figure}[H]
\begin{center}
  \scalebox{0.45}{\includegraphics{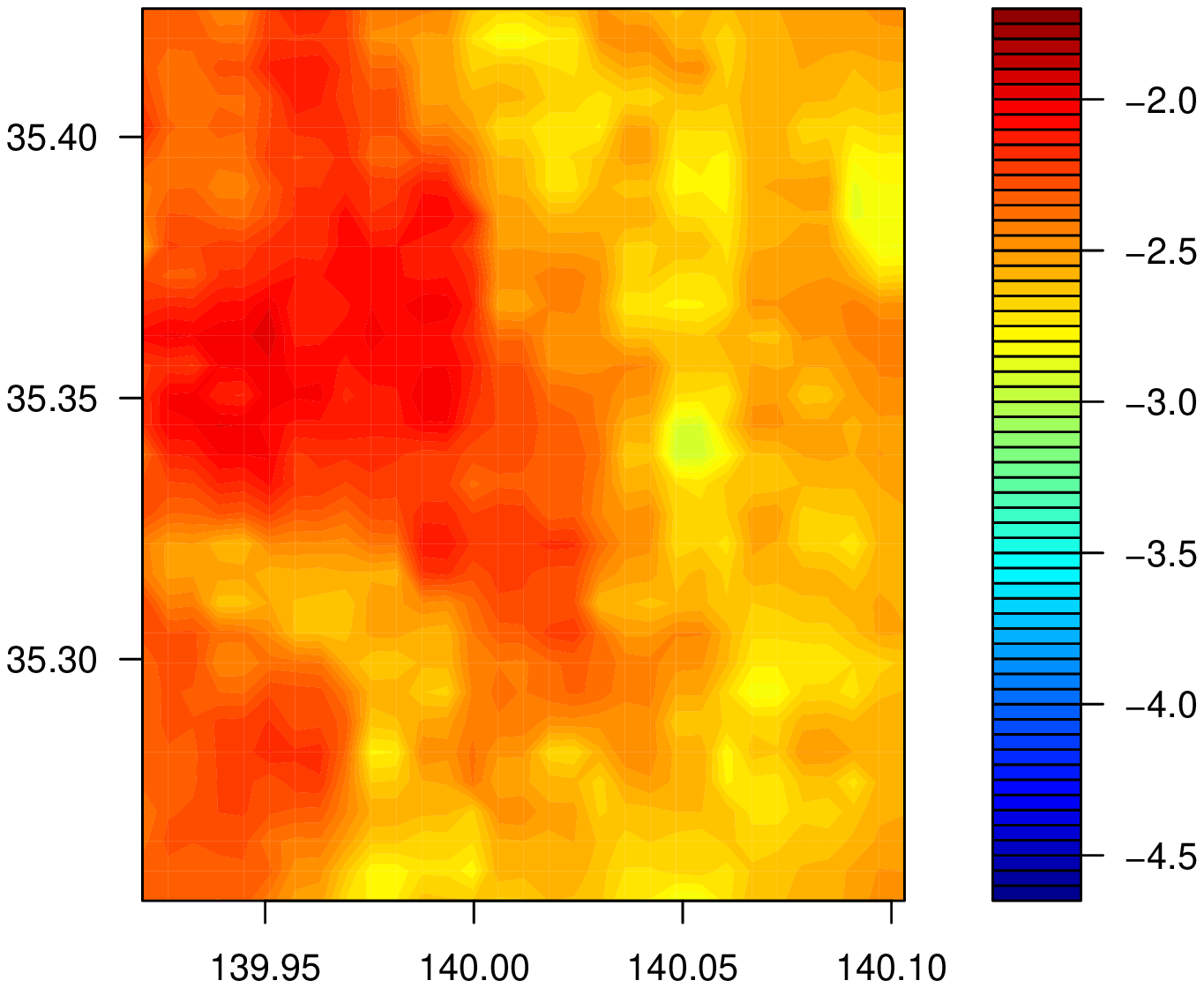}}
  \scalebox{0.45}{\includegraphics{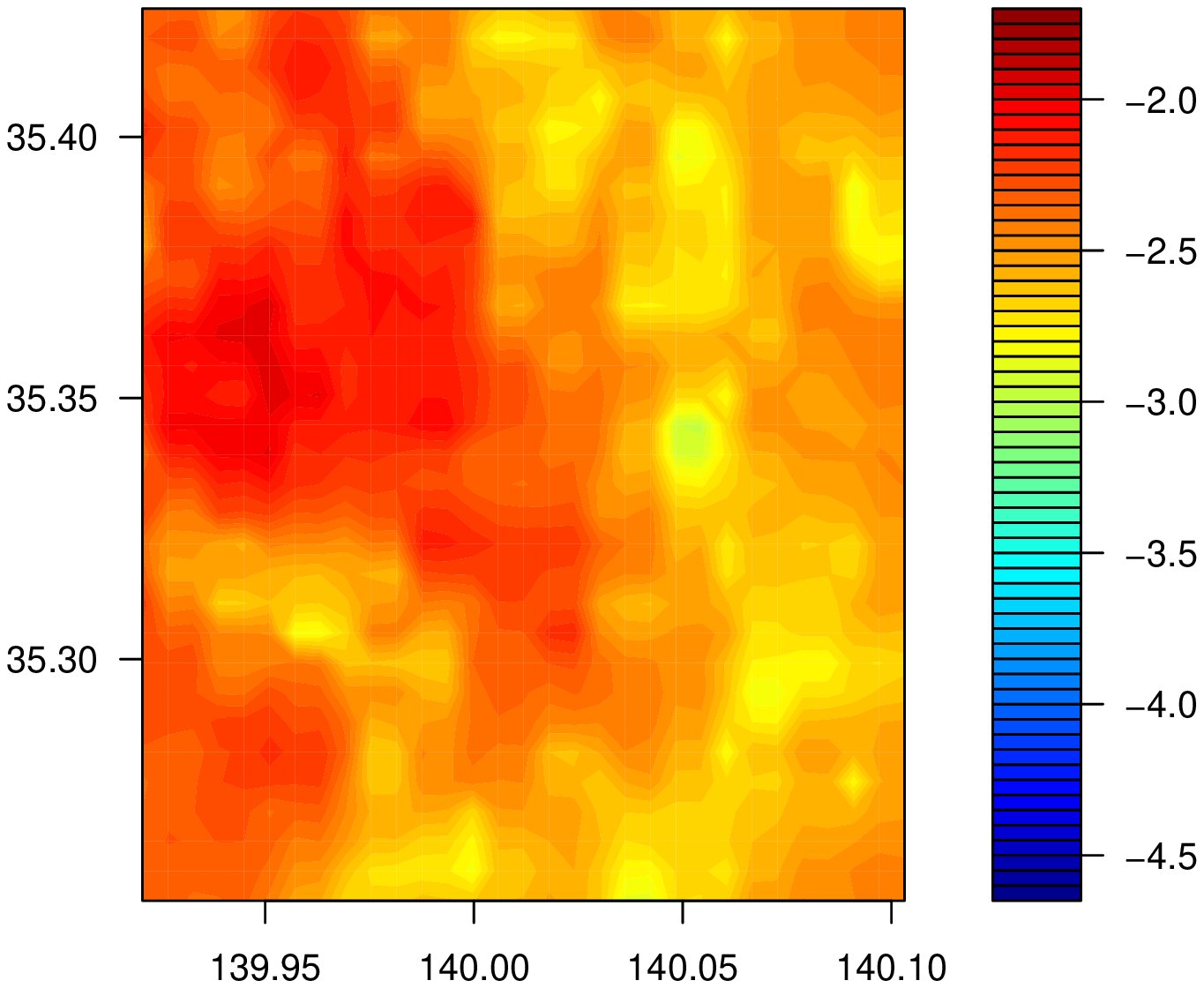}}
  \scalebox{0.45}{\includegraphics{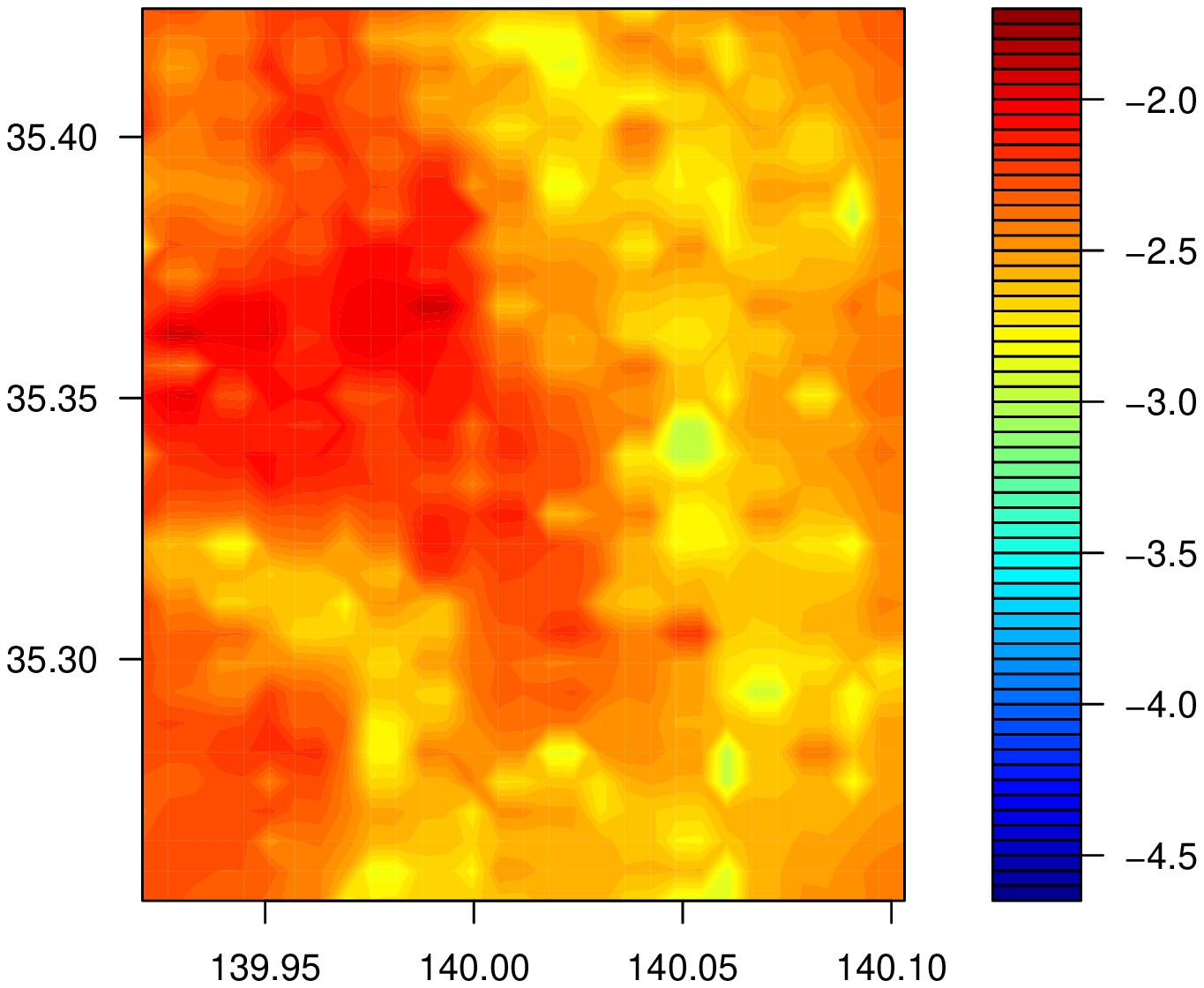}}
  \scalebox{0.45}{\includegraphics{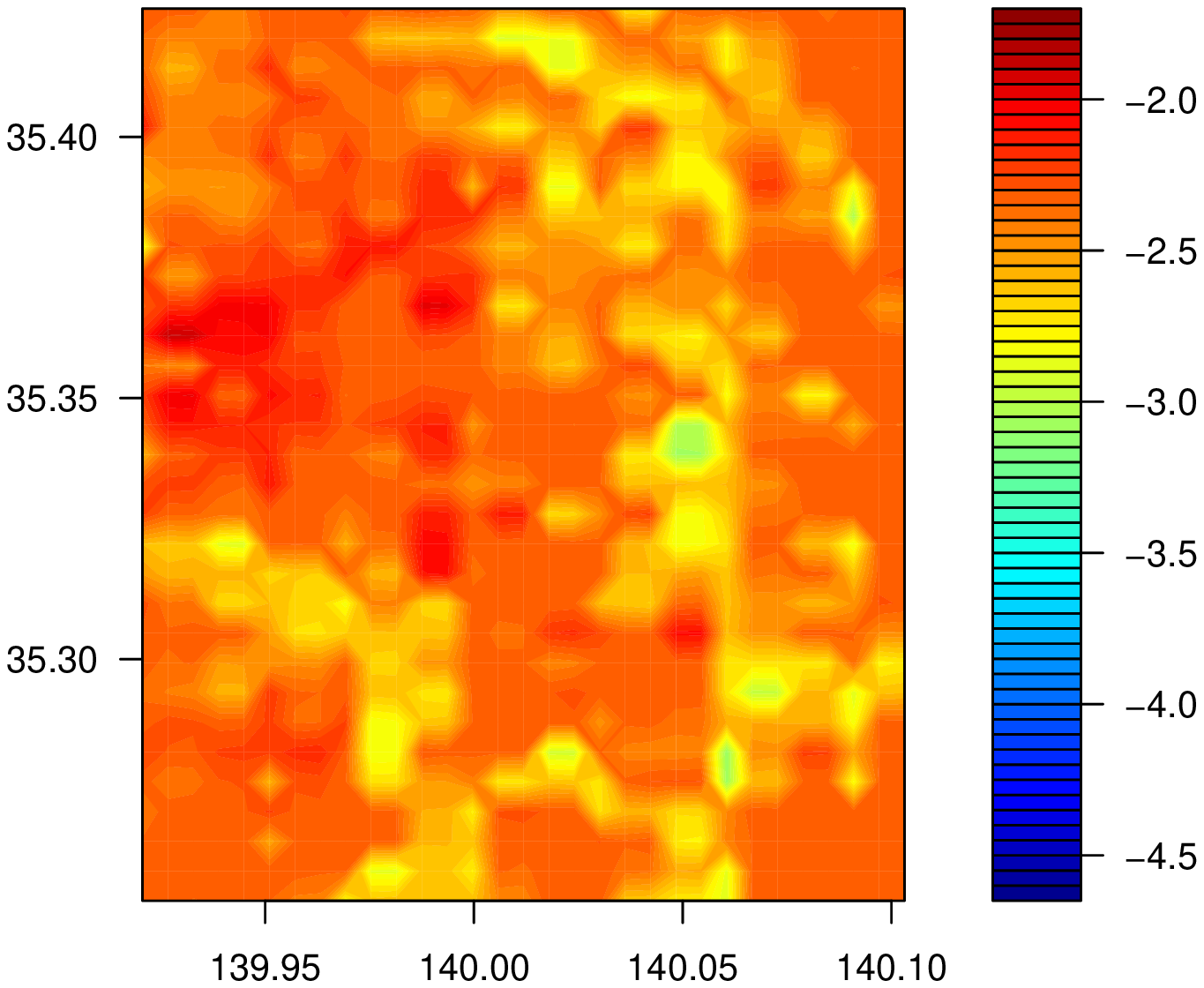}}
  \caption{95\%-quantile of the predictive distribution. We adopt the stationary model and the linear projection for the left and right column in the top row respectively. The modified linear projection and the covariance tapering are used for the left and right column in the bottom row respectively.}
  \label{topsurface}
\end{center}
\end{figure}

\begin{figure}[H]
\begin{center}
  \scalebox{0.45}{\includegraphics{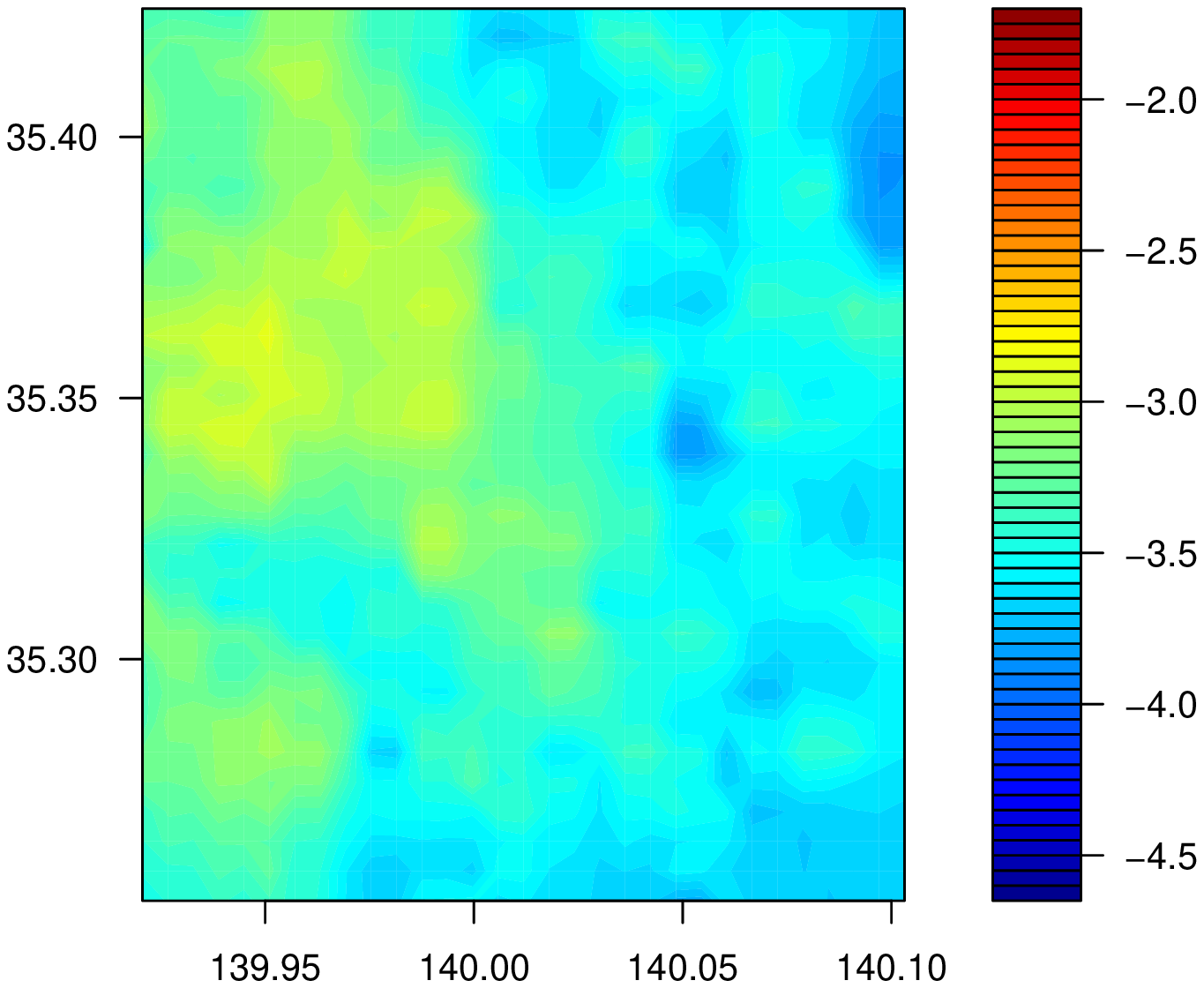}}
  \scalebox{0.45}{\includegraphics{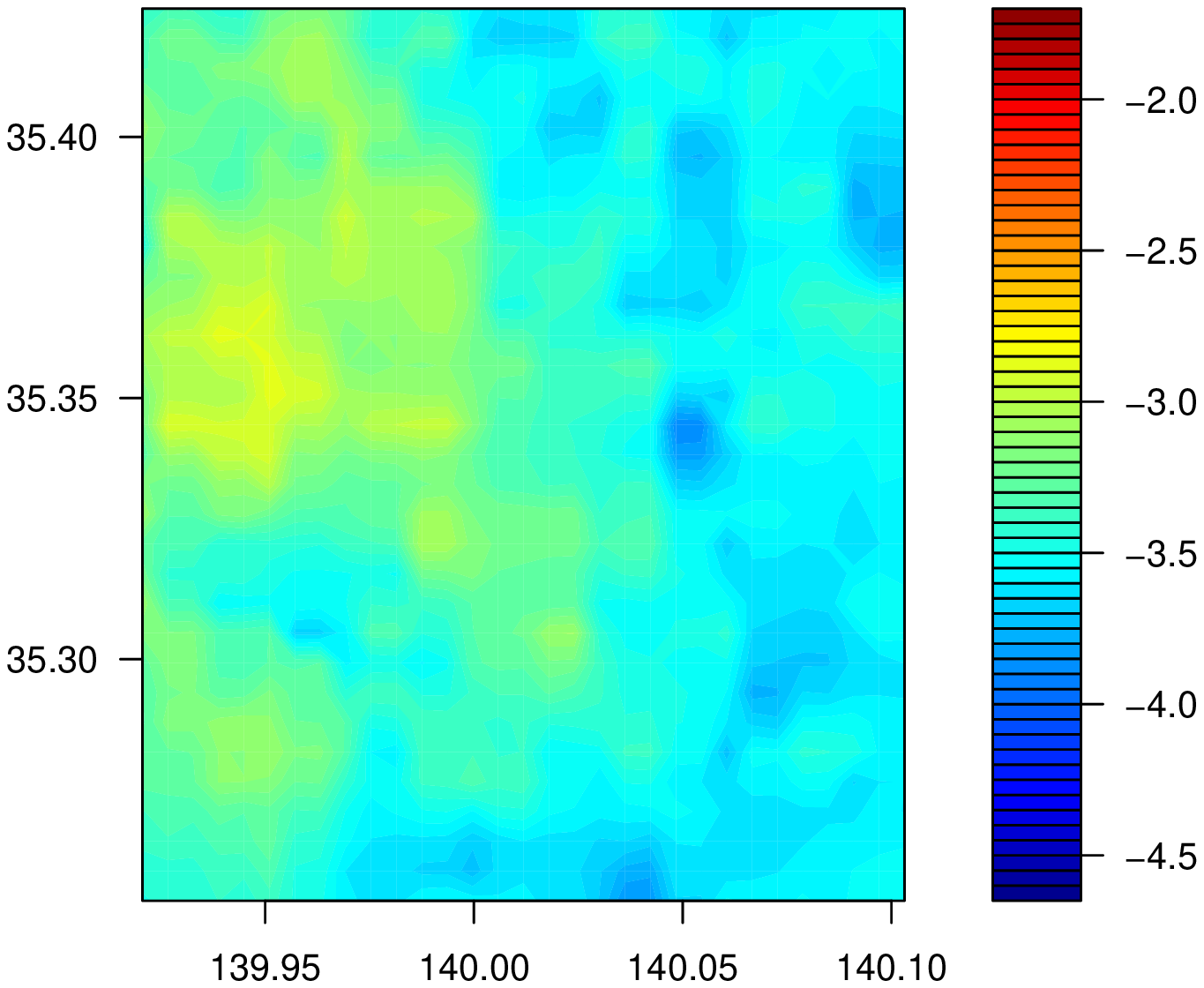}}
  \scalebox{0.45}{\includegraphics{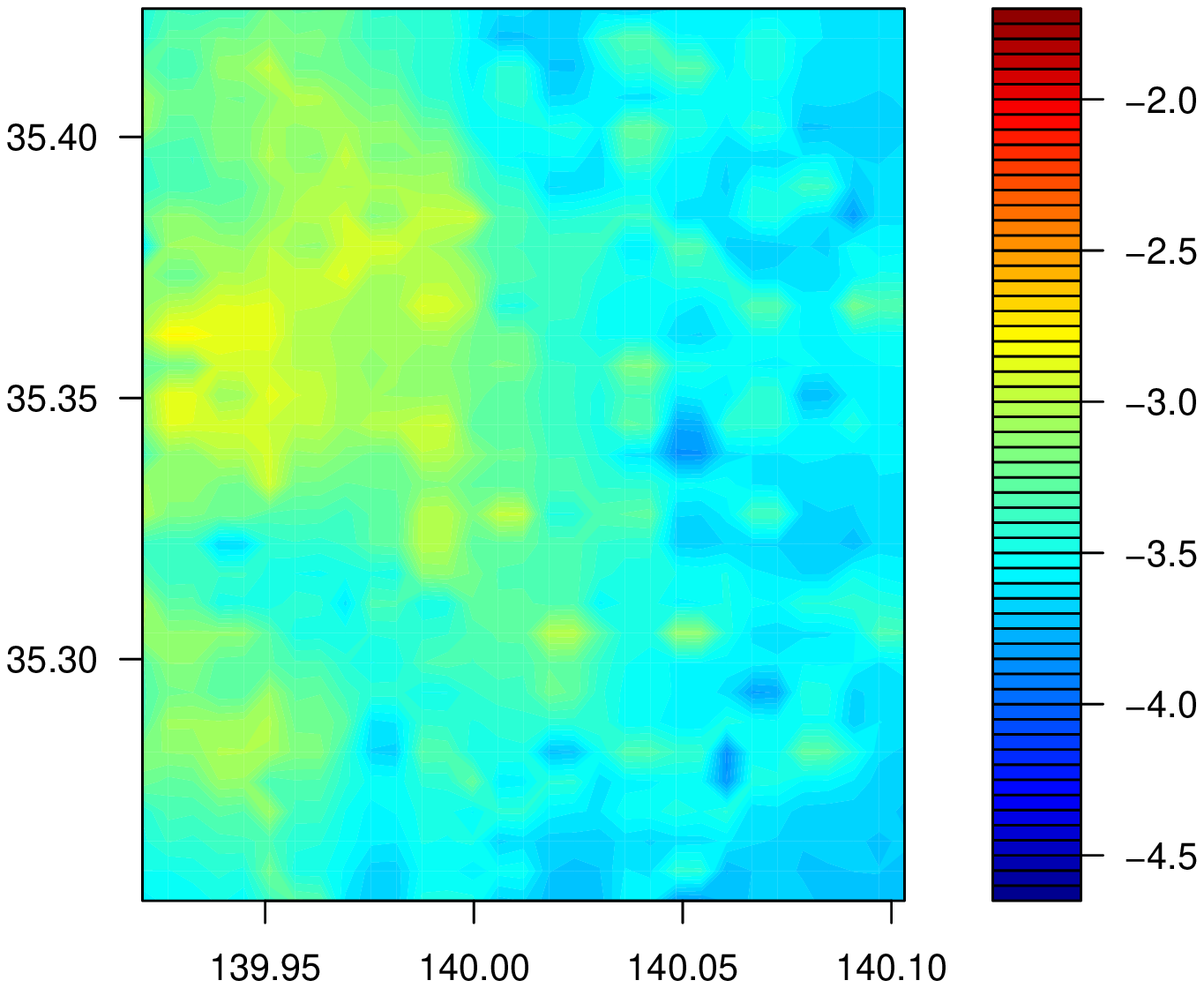}}
  \scalebox{0.45}{\includegraphics{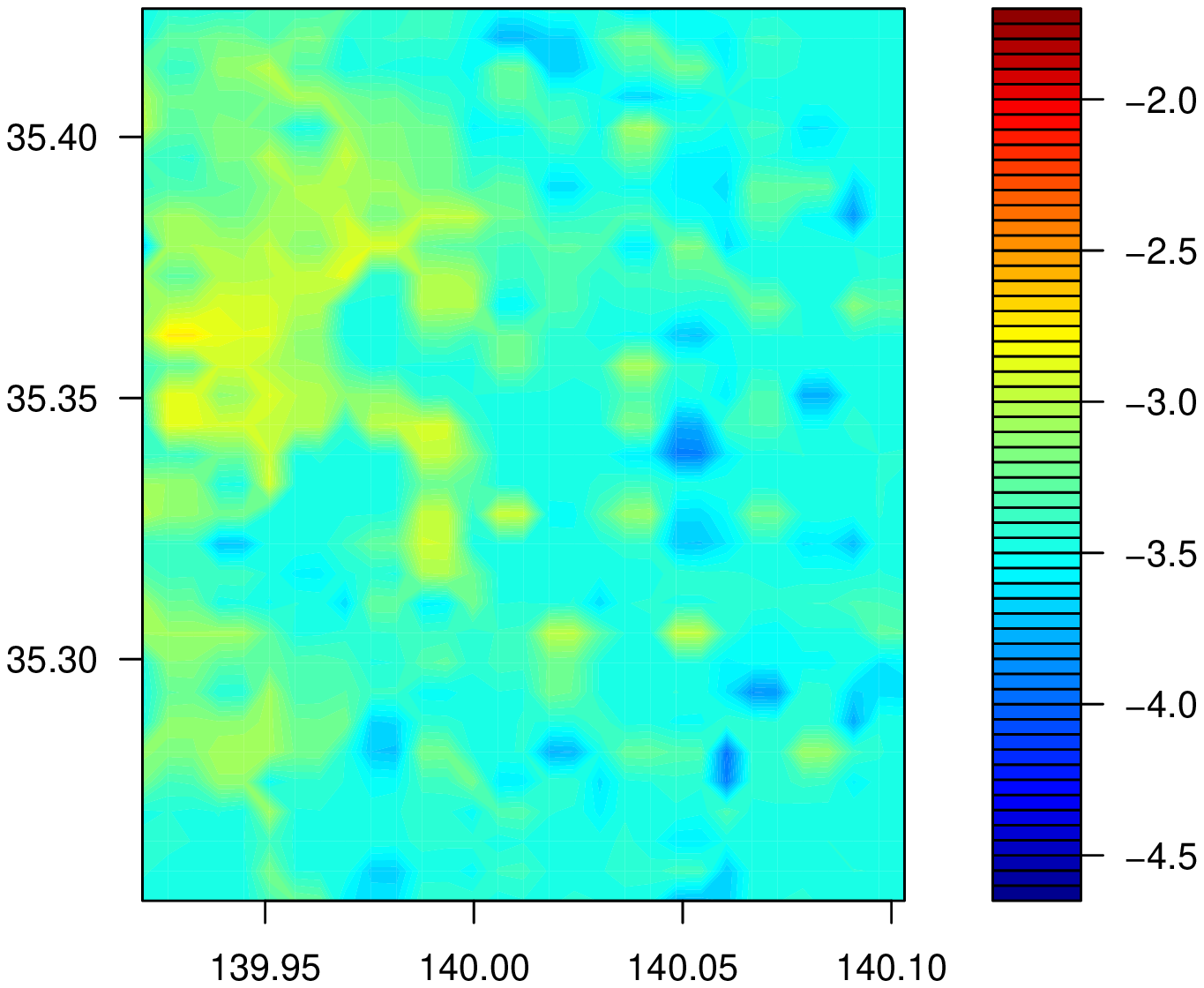}}
  \caption{5\%-quantile of the predictive distribution. We adopt the stationary model and the linear projection for the left and right column in the top row respectively. The modified linear projection and the covariance tapering are used for the left and right column in the bottom row respectively.}
  \label{bottomsurface}
\end{center}
\end{figure}

The result of the Bayesian analysis is shown in Table \ref{empirical}. The modified linear projection has the lowest MSPE and DIC and outperforms even the stationary model with the exponential covariance function which induces the linear projection and modified one. This would suggest that the original spatial data set shows nonstationarity. In the four cases, the estimate of $\tau^2$ has the relatively high value compared to the one found in past empirical studies because of the large local variability which is often observed on the east side of Figure \ref{originaldata}. Since the 95\% credible intervals for $\beta_2$ in the four cases include zero, it seems that we cannot detect strong evidence of the effect of the Fukushima Dai-ichi NPP under the settings of this paper. This may be because the  sampling points are not close to the Fukushima Dai-ichi NPP and the survey was conducted less than 2 years after the TEPCO Fukushima Dai-ichi NPP Accident. A widespread sampling domain and another valid model should be used to investigate the influence of the Fukushima Dai-ichi NPP correctly.

Figure \ref{meansurface} is influenced by the estimate of $\lambda$ from each method. The small value of $\lambda$ in the linear projection results in the smooth surface. In contrast, the predictive surface of the covariance tapering has some small clusters due to the high value of $\lambda$ and large local variability. The modified linear projection shares the features of both the linear projection and the covariance tapering. The original stationary model has a  value of $\lambda$ that is similar to the modified linear projection, but its prediction surface does not express nonstationarity. Figures \ref{topsurface} and \ref{bottomsurface} show large variations of the interquartile ranges at $31 \times 31$ prediction points on the east side of the sampling domain because the original data include large local variability in that region.

It is evident that our proposed modification of the linear projection using the compactly supported correlation function improves the Bayesian analysis more effectively than increasing $\epsilon$. In addition, the modified linear projection serves as a kind of nonstationary covariance function.

\section{Conclusion and future studies}

In this paper, we have proposed a modified linear projection approach for huge irregularly spaced data analysis. Through some simulations and the empirical study, the performance of the linear projection and covariance tapering depends on the dependence properties of the spatial covariance functions. On the other hand, our proposed method is easy to implement and is generally efficient in terms of computation time, estimation of model parameters, and prediction at 
unobserved locations because it effectively captures both the large- and small-scale spatial variations. Moreover, although the modified linear projection was motivated
 by improving the approximation of the original covariance function in the linear projection, the empirical study has shown that it can also be used as a nonstationary covariance function instead of just an approximation.

In the empirical data example, we chose the target error $\epsilon$ and taper range $\gamma$ in consideration of the trade-off between prediction accuracy and computational cost for a subset of the data. In the future, we intend to  develop a comprehensive selection method for these two parameters. It will also be interesting to extend the current work to non-Gaussian, multivariate, and spatio-temporal processes.
%
%
\par
\bigskip
\noindent
\textbf{Appendix : Proofs of Propositions 1 and 2}
\bigskip
\par
\noindent
\textbf{Proof of Proposition 1}
\bigskip

(a) 

Consider $\bm{a}^{\prime} \Phi \Sigma_W \Phi^{\prime} \bm{a}$ for any $\bm{a} \in \mathbb{R}^n / \{ \bm{0} \}$. Setting $\bm{b} = \Phi^{\prime} \bm{a}$, $\bm{b} \neq \bm{0}$ because $\Phi$ is the full row-rank matrix and $\bm{a} \neq \bm{0}$. Thus, $\bm{b}^{\prime} \Sigma_W \bm{b} > 0$.

(b) 

Let a lower triangular matrix $L$ be a Cholesky factor of $\Sigma_W$, that is $\Sigma_W = L L^{\prime}$. Now, we have 
\[
\left\{ \Sigma_{W} - \Sigma_{W} \Phi^{\prime} \left( \Phi \Sigma_{W} \Phi^{\prime}  \right)^{-1} \Phi \Sigma_{W} \right\} = L \left[ \bm{I} - \left(\Phi L \right)^{\prime}  \left\{ \Phi L \left(\Phi L \right)^{\prime}  \right\}^{-1} \Phi L \right] L^{\prime}.
\]
Since $L$ is nonsingular and $\left[ \bm{I} -  \left(\Phi L \right)^{\prime} \left\{ \Phi L \left(\Phi L \right)^{\prime}  \right\}^{-1} \Phi L \right]$ is a projection matrix from $rank \left( \Phi L \right) = m$, $\left\{ \Sigma_{W} - \Sigma_{W} \Phi^{\prime} \left( \Phi \Sigma_{W} \Phi^{\prime}  \right)^{-1} \Phi \Sigma_{W} \right\}$ is positive semidefinite. From Theorem 5.2.1 of Horn and Johnson (1991; page 309), $ \left\{ \Sigma_{W} - \Sigma_{W} \Phi^{\prime} 
\left( \Phi \Sigma_{W} \Phi^{\prime}  \right)^{-1} \Phi \Sigma_{W} \right\}  \circ \Sigma_{taper}$ is also positive semidefinite. Therefore, $\Sigma_{sparse}+ \tau^2 \bm{I} =  \left\{ \Sigma_{W} - \Sigma_{W} \Phi^{\prime} 
\left( \Phi \Sigma_{W} \Phi^{\prime}  \right)^{-1} \Phi \Sigma_{W} \right\}  \circ \Sigma_{taper}  + \tau^2 \bm{I}$ is positive definite.

(c) 

From (a), $\Sigma_{W} \Phi^{\prime} \left( \Phi \Sigma_{W} \Phi^{\prime}  \right)^{-1} \Phi \Sigma_{W}$ is positive semidefinite. \\
Thus, $\Sigma_{mlp} + \tau^2 \bm{I} =  \Sigma_{W} \Phi^{\prime} 
\left( \Phi \Sigma_{W} \Phi^{\prime}  \right)^{-1} \Phi \Sigma_{W} + \Sigma_{sparse}  + \tau^2 \bm{I}$ is positive definite.

(d) 

It is clear from (a) and (b).

\hfill $\Box$

\bigskip
\par
\noindent
\textbf{Proof of Proposition 2}
\bigskip

In this proof, we denote $(\Sigma_W)_{ij} = a_{ij}$, $(\Sigma_{approx})_{ij} = b_{ij}$, $(\Sigma_{taper,\gamma_1})_{ij} = d^{(1)}_{ij}$, $(\Sigma_{taper,\gamma_2})_{ij} = d^{(2)}_{ij}$ and $(\bm{I})_{ij} = e_{ij}$ for $i,j = 1,\ldots,n$. It follows that 
\begin{align*}
\| \Sigma_W - \Sigma_{approx} \|_F &= \sum_{i=1}^{n}\sum_{j=1}^{n} (a_{ij} - b_{ij})^2, \\
\| \Sigma_W - \left\{ \Sigma_{approx} + \left( \Sigma_W - \Sigma_{approx} \right) \circ \bm{I}  \right\}  \|_F &= 
 \sum_{i=1}^{n}\sum_{j=1}^{n} (a_{ij} - b_{ij})^2 (1 - e_{ij})^2, \\
 \intertext{and}
\| \Sigma_W - \left\{ \Sigma_{approx} + \left( \Sigma_W - \Sigma_{approx} \right) \circ \Sigma_{taper,\gamma_k}  \right\}  \|_F &=  \sum_{i=1}^{n}\sum_{j=1}^{n} (a_{ij} - b_{ij})^2 \left(1 - d^{(k)}_{ij} \right)^2, 
\end{align*}
for $k = 1,2$. From $1 \ge (1-e_{ij})^2 \ge \left(1 - d^{(1)}_{ij} \right)^2 \ge \left(1 - d^{(2)}_{ij} \right)^2$, the proof is completed.

\hfill $\Box$
%
%
\par
\bigskip
\noindent
\textbf{Acknowledgments} The author is grateful to Professor Yasuhiro Omori and Professor Yoshihiro Yajima for their  helpful comments and discussions. I also acknowledge the suggestions from the editor and the anonymous referee that refined and improved the manuscript. All results remain the author's responsibility.
%
%

\end{document}